\documentclass[10pt]{article}
\usepackage{amsmath}
\usepackage{amssymb}
\usepackage{pgf}
\usepackage{pgfplots}
\pgfplotsset{width=2.4in, height=1.9in,every axis plot/.append style={thick}, compat=1.9}
\usepackage{graphicx}
\usepackage{epstopdf, epsfig}
\usepackage[left=2.50cm, right=2.50cm, top=3.00cm, bottom=3.00cm]{geometry}
\usepackage{ctable}
\graphicspath{{./fig/}}

\newcommand{\Rey}{\mbox{\textit{Re}}}

\newcommand{\Pran}{\mbox{\textit{Pr}}}

\newcommand{\avg}[2]{\langle#1\rangle_{\mathrm{#2}}}
\newcommand{\vtavg}[1]{\langle#1\rangle}

\renewcommand{\vec}[1]{\mathbf{#1}}
\newcommand{\citep}[1]{\cite{#1}}
\newcommand{\citet}[1]{\cite{#1}}

\newcommand\Nu{\mbox{\textit{Nu}}}
\newcommand\Ra{\mbox{\textit{Ra}}}
\newcommand\Gr{\mbox{\textit{Gr}}}
\newcommand{\aff}[1]{$^#1$}

\begin{document}

\begin{center}
{\LARGE\bf Inclined turbulent thermal convection in liquid sodium}\\[1em]

{Lukas Zwirner\aff{1}, Ruslan Khalilov\aff{2}, Ilya Kolesnichenko\aff{2},\\
Andrey Mamykin\aff{2}, Sergei Mandrykin\aff{2}, Alexander Pavlinov\aff{2}, Alexander Shestakov\aff{2}, Andrei Teimurazov\aff{2},\\
Peter Frick\aff{2} \& Olga Shishkina\aff{1}}\\[2em]
{\small \aff{1}Max Planck Institute for Dynamics and Self-Organization,\\ Am Fassberg 17, 37077 G\"ottingen, Germany\\
\aff{2}Institute of Continuous Media Mechanics,\\ Korolyov 1, Perm, 614013, Russia}
\end{center}
\begin{abstract}
Inclined turbulent thermal convection by large Rayleigh numbers
in extremely small-Prandtl-number fluids is studied based on results of both, measurements and high-resolution numerical simulations.
The Prandtl number $\Pran\approx0.0093$ considered in the experiments and the Large-Eddy Simulations (LES) and 
$\Pran=0.0094$ considered in the Direct Numerical Simulations (DNS)
correspond to liquid sodium, which is used in the experiments.
Also similar are the studied Rayleigh numbers, which are, respectively, $\Ra=1.67\times10^7$ in the DNS,
$\Ra=1.5\times10^7$ in the LES and $\Ra=1.42\times10^7$ in the measurements.
The working convection cell is a cylinder with equal height and diameter, where one circular surface is heated and another one is cooled.
The cylinder axis is inclined with respect to the vertical and the inclination angle varies from $\beta=0^\circ$,
which corresponds to a Rayleigh--B\'enard configuration (RBC), to $\beta=90^\circ$, as in a vertical convection (VC) setup.
The turbulent heat and momentum transport as well as time-averaged and instantaneous flow structures and
their evolution in time are studied in detail, for different inclination angles, and are illustrated also by supplementary videos,
obtained from the DNS and experimental data.

To investigate the scaling relations of the mean heat and momentum transport in the limiting cases of RBC and VC configurations,
additional measurements are conducted for about one decade of the Rayleigh numbers around $\Ra=10^7$ and $\Pran\approx0.009$.
With respect to the turbulent heat transport in inclined thermal convection by low $\Pran$,
a similarity of the global flow characteristics for the same value of $\Ra\Pran$ is proposed and analysed, based on the above simulations
and measurements and on complementary DNS for $\Ra=1.67\times10^6$, $\Pran=0.094$ and $\Ra=10^9$, $\Pran=1$.
\end{abstract}


\section{Introduction}

Elucidation of the mechanisms of turbulent thermal convection in very-low-Prandtl fluids that takes place, for example, on surfaces of stars, 
including the Sun, where the Prandtl number ($\Pran$) varies from $10^{-8}$ to $10^{-4}$ \citep{Spiegel1962, Hanasoge2016}, is crucial for understanding of the universe.
One of the possible ways to get one step closer to this goal is to investigate laboratory convective flows,
which can be classified as turbulent and which are characterised by very small Prandtl numbers ($\Pran<10^{-2}$).
In the present experimental and numerical study we focus on the investigation of turbulent natural thermal convection in liquid sodium ($\Pran\approx0.0093$), 
where the imposed temperature gradient, like in nature and in many engineering applications,  is not necessarily parallel to the gravity vector.

Generally, a turbulent fluid motion, which is driven by an imposed temperature gradient, is a very common phenomenon in nature and is important in many industrial applications.
In one of the classical models of thermal convection, which is Rayleigh--B\'enard convection (RBC),
the fluid is confined between a heated lower horizontal plate and an upper cooled plate,
and buoyancy is the main driving force there: The temperature inhomogeneity leads to the fluid density variation, which in presence of gravity leads to a convective fluid motion.
For reviews on RBC we refer to \cite{Bodenschatz2000, Ahlers2009, Lohse2010, Chilla2012}.

In the case of convection under a horizontal temperature gradient, which is known as vertical convection (VC) or convection in cavities, the heated and cooled plates are parallel to each other, 
as in RBC, but are located parallel to the gravity vector. 
Therefore, in this case shear plays the key role, see \cite{Ng2015, Ng2017, Shishkina2016c}.
The concept of inclined convection (IC) is a generalisation of RBC and VC.
There, the fluid layer, heated on one surface and cooled from the opposite surface, is tilted with respect to the gravity direction, so that
both, buoyancy and shear drive the flow in this case.
This type of convection was studied previously, in particular, by
\cite{Daniels2003, Chilla2004, Sun2005b, Ahlers2006a, Riedinger2013, Weiss2013, Langebach2014}
and more recently by \cite{Frick2015, Mamykin2015, Vasiliev2015, Kolesnichenko2015, Shishkina2016b, Teimurazov2017, Mandrykin2018, Khalilov2018, Zwirner2018}.

In thermal convection, the global flow structures and heat and momentum transport are determined mainly by the following system parameters:
the Rayleigh number $\Ra$, Prandtl number $\Pran$ and the aspect ratio of the container $\Gamma$:
\begin{eqnarray}
\label{Ra}
\Ra\equiv \alpha g \Delta L^3/(\kappa \nu),\\
\quad\Pran\equiv \nu/\kappa, \quad \Gamma\equiv D/L.\nonumber
\end{eqnarray}
Here $\alpha$ denotes the isobaric thermal expansion coefficient,
$\nu$ the kinematic viscosity,
$\kappa$ the thermal diffusivity of the fluid,
$g$ the acceleration due to gravity,
$\Delta\equiv T_+-T_-$ the difference between the temperatures at the heated plate ($T_+$) and at the cooled plate ($T_-$),
$L$ the distance between the plates
and $D$ the diameter of the plates.

The main response characteristics of a natural convective system are
the mean total heat flux across the heated/cooled plates, $q$, normalised by the conductive part of the total heat flux, $\hat{q}$,
i.e. the Nusselt number $\Nu$,
and the Reynolds number $\Rey$,
\begin{eqnarray}
 \Nu\equiv q/\hat{q}, \qquad \Rey\equiv LU/\nu.
\end{eqnarray}
Here $U$ is the reference velocity,
which is usually determined by either the maximum of the time-averaged velocity along the plates or by ${\langle\text{\bf u}\cdot\text{\bf u}\rangle^{1/2}}$,
i.e. it is based on the mean kinetic energy,
with $\text{\bf u}$ being the velocity vector-field and
$\langle \cdot\rangle$ denotes the average in time and over the whole convection cell.
Note, that even for a fixed setup in natural thermal convection, where no additional shear is imposed into the system, the scaling relations of the mean heat and momentum transport, represented by $\Nu$ and $\Rey$, with the input parameters $\Ra$ and $\Pran$,
are not universal and are influenced by non-Oberbeck--Boussinesq (NOB) effects, see \cite{Kraichnan1962, Grossmann2000, Ahlers2006, Ahlers2009, Lohse2010, Shishkina2016a, Shishkina2016d, Weiss2018}.

Here one should note that apart from $\Pran$ and $\Ra$, the geometrical confinement of the convection cell also determines the strength of the heat transport
\citep{Huang2013, Chong2015, Chong2016}.
Thus, in experiments by \cite{Huang2013} for $\Pran=4.38$, an increase of $\Nu$ due to the cell confinement was obtained, while in the
Direct Numerical Simulations (DNS) by \cite{Wagner2013} for $\Pran = 0.786$,
the heat and mass transport gradually reduced with increasing confinement.
This virtual contradiction was recently resolved in \cite{Chong2018}.
It was found that $\Pran$ determines whether the optimal $\Gamma$,
at which the maximal heat transport takes place, exists or not.
For $\Pran > 0.5$ ($\Ra = 10^8$)  an enhancement of $\Nu$ was observed, where the optimal $\Gamma$ decreases with increasing $\Pran$, but for $\Pran\leq0.5$ a gradual reduction of the heat transport with increasing confinement was obtained.
For all $\Pran$, the confinement induced friction causes a reduction of $\Rey$.

In a general case of inclined thermal convection, apart from $\Ra$, $\Pran$ and the geometry of the container,
also the cell inclination angle $\beta$ ($\beta=0^\circ$ in the RBC configuration and $\beta=90^\circ$ in VC)
is the influential input parameter of the convective system.
Experimental studies of turbulent thermal liquid sodium convection in cylinders of different aspect ratios,
showed that the convective heat transfer between the heated and cooled
parallel surfaces of the container is most efficient neither in a standing position of the cylinder
(as in RBC, with a cell inclination angle $\beta = 0^\circ$), nor in a lying position (as in VC, $\beta = 90^\circ$),
but in an inclined position for a certain intermediate value of $\beta$, $0^\circ < \beta < 90^\circ$,
see \cite{Vasiliev2015} for $L=20D$, \cite{Mamykin2015} for $L=5D$ and \cite{Khalilov2018} for $L=D$.
Moreover, these experiments showed that for $\Pran\ll1$ and $\Ra\gtrsim10^9$, any tilt $\beta$,
$0^\circ<\beta\leq90^\circ$, of the cell leads to a larger mean heat flux ($\Nu$) than in RBC,
by similar values of $\Ra$ and $\Pran$.
Note that the effect of the cell tilting on the convective heat transport in low-$\Pran$ fluids is very different from that in the case of large $\Pran$ \citep{Shishkina2016b}.
For example, for $\Pran\approx6.7$ and $\Ra\approx4.4\times10^9$, a monotone reduction of $\Nu$
with increasing $\beta$ in the interval $\beta\in[0^\circ,90^\circ]$ takes place, as it was obtained in
measurements by \cite{Guo2015}.

One should mention that there are only a few experimental and numerical studies of IC in a broad range of $\beta$,
whereas most of the investigations of the cell-tilt effects on the mean heat transport
were conducted in a narrow region of $\beta$ close to $0^\circ$ and mainly for large-$\Pran$ fluids.
These studies showed generally a small effect of $\beta$ on $\Nu$,
reflected in a tiny reduction of $\Nu$ with increasing $\beta$ close to $\beta=0^\circ$,
see \cite{Ciliberto1996, Cioni1997, Chilla2004, Sun2005b, Ahlers2006a, Roche2010, Wei2013}.
A tiny local increase of $\Nu$ with a small inclination of the RBC cell filled with a fluid of $\Pran>1$ is possible only
when a two-roll form of the global Large Scale Circulation (LSC) is present in RBC,
which usually almost immediately transforms into a single-roll form of the LSC with any inclination \citep{Weiss2013}.
The single-roll LSC is known to be more efficient in the heat transport than its double-roll form,
as it was proved in the measurements \citep{Xi2008, Weiss2013} and DNS \citep{Zwirner2018}.
Thus, all available experimental and numerical results on IC show that
the $\Nu(\beta)/\Nu(0)$ dependence is a complex function of $\Ra$, $\Pran$ and $\Gamma$,
which cannot be represented as a simple combination of their power functions.

A strong analogy can be seen between the IC flows and convective flows, which occur from the imposed temperature differences
at both, the horizontal and vertical surfaces of a cubical container.
With a different balance between the imposed horizontal and vertical temperature gradients, 
where the resulting effective temperature gradient has non-vanishing horizontal and vertical components,
one can mimic the IC flows by different inclination angles. 
Experimental studies on these type of convective flows were conducted by \cite{Zimin1982}.

Although there is no scaling theory for general IC,
for the limiting configurations in IC ($\beta=0^\circ$ and $\beta=90^\circ$) and sufficiently wide heated/cooled plates,
there are theoretical studies of the scaling relations
of $\Nu$ and $\Rey$ with $\Pran$ and $\Ra$.
For RBC ($\beta=0$), \cite{Grossmann2000, Grossmann2001, Grossmann2011} (GL) developed a scaling theory
which is based on a decomposition into boundary-layer (BL) and bulk contributions
of the time- and volume-averaged kinetic ($\epsilon_u$) and thermal ($\epsilon_\theta$) dissipation rates,
for which there exist analytical relations with $\Nu$, $\Ra$ and $\Pran$.
Equating $\epsilon_u$ and $\epsilon_\theta$ to their estimated either bulk or BL contributions
and employing in the BL dominated regimes the Prandtl--Blasius BL theory
\citep{Prandtl1905, Blasius1908, Landau1987, Schlichting2000}, theoretically possible limiting scaling regimes were derived.
The theory allows to predict $\Nu$ and $\Rey$ in RBC
if the pre-factors  fitted with the latest experimental and numerical data are used, see \cite{Stevens2013} and \cite{Shishkina2017}.

In contrast to RBC, in the other limiting case of IC, which is VC ($\beta=90^\circ$),
the mean kinetic dissipation rate $\epsilon_u$ cannot be derived analytically from $\Ra$, $\Pran$ and $\Nu$, as in RBC,
and this impedes an extension of the GL theory to VC.
However, for the case of laminar free convection between two differently heated plates (i.e., VC),
it is possible to derive the dependences of $\Rey$ and $\Nu$ on $\Ra$ and $\Pran$
from the BL equations, under an assumption that a similarity solution exists \citep{Shishkina2016c}.
Although this problem is solved for the laminar case,
to our knowledge, there is no theoretical model to predict $\Nu$ and $\Rey$ in turbulent VC.
It is expected, however that in the asymptotic regime by high $\Ra$,
the scaling exponents in the $\Nu$ vs. $\Ra$ and $\Rey$ vs. $\Ra$ scalings is 1/2, as in RBC \citep{Ng2018}.

Generally, the dependences of $\Nu$ and $\Rey$ on $\Ra$ and $\Pran$ in VC have been less investigated than those in RBC.
For similar cell geometry and ranges of $\Ra$ and $\Pran$, not only the heat transport in VC differs quantitatively from that in RBC
\citep{Bailon2012, Wagner2013, Wagner2015, Ng2015}, but the VC and RBC flows can be even in different states.
For example, for $\Pran=1$, $\Ra=10^8$ and a cylindrical container of $\Gamma=1$, the VC flow is steady, while the RBC flow is already turbulent
\citep{Shishkina2016b}.
Previous experimental and numerical studies of free thermal convection under an imposed horizontal temperature gradient (i.e., VC)
reported the scaling exponent $\gamma$ in the power law $\Nu\sim\Ra^\gamma$, varying from 1/4 to 1/3.
In laminar VC, it is about 1/4 \citep{Schmidt1930, Lorenz1934, Saunders1939, Churchill1975},
being slightly larger for very small $\Ra$, where the geometrical cell confinement influences the heat transport
\citep{Versteegh1999, Yu2007, Kis2012, Ng2015}.
The scaling exponent $\gamma$ is also larger for very high $\Ra$, where,
with growing $\Ra$, the VC flows become first transitional \citep{Ng2017} and
later on fully turbulent \citep{George1979, Fujii1970}.
Note that all the mentioned experiments and simulations of VC were conducted for fluids of $\Pran$ about or larger than 1.

For the case of small $\Pran$, in the experiments by \cite{Frick2015} and \citet{Mamykin2015}
on turbulent VC in liquid sodium ($\Pran\approx 0.01$) in elongated cylinders,
significantly larger scaling exponents were observed, due to the geometrical confinement.
Thus, for a cylinder with $L=5D$ and the Rayleigh numbers, based on the cylinder diameter,  up to $10^7$,
\cite{Frick2015} obtained $\Nu \sim \Ra^{0.43}$ and $\Rey \sim \Gr^{0.44}$, where
$\Gr$ is the Grashof number, $\Gr\equiv \Ra/\Pran$.
For an extremely strong geometrical confinement, namely, for a cylindrical convection cell with $L=20D$,
and  a similar Rayleigh number range, \citet{Mamykin2015}  found
$\Nu  \sim  \Ra^{0.95}$ and  $\Rey \sim \Gr^{0.63}$.

As already mentioned, natural thermal convection for $\Pran\ll1$ is significantly less studied than that for $\Pran\sim1$,
whereas the knowledge of convection in very-low-Prandtl-number fluids is desired for understanding of many astrophysical phenomena and
needed in engineering and technology.
Thus, liquid metals, being very efficient heat transfer fluids, are used in numerous applications.
Of particular interest is liquid sodium, which is one of the liquid metals that has been mostly wide used in liquid metal-cooled fast neutron reactors being
developed during the last several decades  \citep{Heinzel2017}.

Although the knowledge on natural thermal convection in liquid metals is required for the development of safe and efficient liquid metal heat exchangers,
the experimental database of the corresponding measurements remains to be quite restricted due to
the known difficulties in conduction of thermal measurements in liquid metals.
Apart from general problems, occurring from the high temperature and aggressivity of liquid sodium, natural convective flows 
are known to be relatively slow and are very sensitive to the imposed disturbances.
While the probe measurements in the core part of the flows are possible in pipe and channel flows \citep{Heinzel2017, Onea2017a, Onea2017},
since the interference influences the flows only downstreams in those cases, they unavoidably induce too strong disturbances in the case of natural thermal convection.

There exist a few measurements of the scaling relations of $\Nu$ vs. $\Ra$ in liquid-metal Rayleigh--B\'enard convection (without any cell inclination).
Thus, for mercury ($\Pran=0.025$), it was reported $\Nu\sim\Ra^{0.27}$ for $2\times10^6 < \Ra < 8\times10^7$ by \cite{Takeshita1996},
$\Nu\sim\Ra^{0.26}$ for $7\times10^6 < \Ra < 4.5\times10^8$
and $\Nu\sim\Ra^{0.20}$ for $4.5\times10^8 < \Ra < 2.1\times10^9$ by \cite{Cioni1997}.
For liquid sodium ($\Pran=0.006$), \cite{Horanyi1999} measured
$\Nu\sim\Ra^{0.25}$ for $2\times10^4 < \Ra < 5\times10^6$.

IC in liquid sodium has been studied so far by \cite{Frick2015, Mamykin2015, Vasiliev2015, Kolesnichenko2015, Khalilov2018}.
These sodium  experiments  were  conducted in relatively long cylinders, in which the scaling exponents
are essentially increased due to the geometrical confinement.
For RBC in a cylinder with $L=5D$, \cite{Frick2015} reported $\Nu \sim \Ra^{0.4}$ and for RBC in a very long
cylinder with  $L=20D$,   \cite{Mamykin2015} obtained   $\Nu \sim \Ra^{0.77}$.
In both studies, also IC in liquid sodium for $\beta=45^\circ$ was investigated,
and the following scaling laws were obtained for this inclination angle:
$\Nu \sim \Ra^{0.54}$ for $L=5D$ and $\Nu \sim \Ra^{0.7}$ for $L=20D$.
Note that in both IC measurements, much higher mean heat fluxes were obtained,
compared to those in VC or RBC configurations.
Liquid sodium IC in a broad range of the inclination angle $\beta$ (from $0^\circ$ to $90^\circ$)
has been studied so far by \citet{Vasiliev2015} for a long cylinder with $L=20D$
and by \cite{Khalilov2018} for a cylinder of the  aspect ratio one ($L=D$).

Conduction of accurate DNS of natural thermal convection by high $\Ra$ and very low $\Pran$ is also very challenging, since
it requires very fine meshes in space and in time, due to the necessity to resolve the Kolmogorov microscales in the bulk of the flows as well in the viscous BLs \citep{Shishkina2010}.
When $\Pran\ll1$, the thermal diffusion, represented by $\kappa$, is much larger than the momentum diffusion, represented by $\nu$,
and therefore, the viscous BLs become extremely thin by large $\Ra$.
Thus, there exist only a few DNS of thermal convection for a combination of large $\Ra$ and very low Prandtl numbers, $\Pran\leq0.025$,
which, moreover, have been conducted exclusively for the Rayleigh--B\'enard configuration, i.e., only for $\beta=0$
\citep{Schumacher2016, Scheel2016, Scheel2017, Horn2017, Vogt2018b}.

In engineering applications, however, of particular interest is natural thermal convection with different orientation in the gravity field of the imposed temperature gradient.
Turbulent thermal convection of liquid metals in such flow configurations is especially relevant
in cooling systems in tokamaks and fast breeder reactors, and therefore
investigation of such flows are of special importance \citep{Zhilin2009, Belyaev2013}.
The above nonferrous-metal devices are known to be strongly affected by turbulent thermal convection.
For example, convective flows of magnesium that develop in the titanium reduction reactors,
are characterised by the Grashof number of order of $10^{12}$.
DNS of such flows by that large $\Gr$ and extremely small Prandtl numbers,
would  be  extremely  expensive and currently are unrealizable and, therefore, further development of
reduced mathematical models is still required.
Computational codes for Large-Eddy Simulations (LES) of turbulent thermal convection in liquid metals,
verified against the corresponding experiments and DNS, can be useful in solving of such kind of problems \citep{Teimurazov2017a}.

Thus, IC measurements in liquid sodium by large $\Ra$ and the corresponding precise numerical simulations are in great demand
and, therefore, we devote our present work to this topic.
By combining experiments, DNS and LES, we are aimed to paint a complementary picture of this kind of convection.
In chapter~2 of this paper, the experiment, DNS, LES and methods of data analysis are described.
Chapter~3 presents the obtained results with respect to the heat and momentum transport in IC in liquid sodium
and the global flow structures and their evolution in time.
An analogy of the global flow structures and global heat transport in flows of similar values of $\Ra\,\Pran$ is also discussed there.
Conclusion chapter summarises the obtained results and gives an outlook on future research.

\section{Methods}

\subsection{Experiment}

\begin{figure}
\centering
\includegraphics{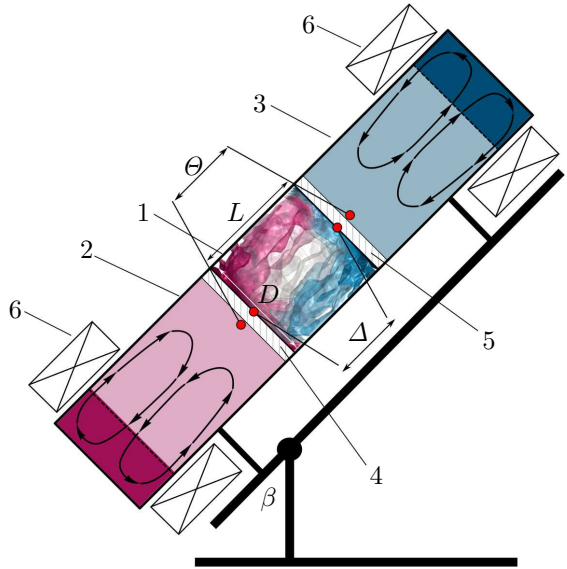}
\caption{Sketch of the experimental facility, which consists of:
(1) a cylindrical convection cell, (2) a hot heat exchanger chamber, (3) a cold heat exchanger chamber,
(4) a heated copper plate, (5) a cooled copper plate and (6) inductor coils.
The convection cell (1) and heat exchangers (2, 3) are filled with liquid sodium.
$D$ is the diameter and $L$ the length of the cylindrical convection cell (1),
$\beta$ is the cell inclination angle,
$\Theta$ the temperature drop between the hot and cold heat exchanger chambers,
$\Delta$ the resulting temperature drop between the inner surfaces of the heated and cooled plates of the convection cell (1). }
\label{PIC1}
\end{figure}

All experimental data presented in this paper are obtained at the experimental facility, described in detail in \citet{Khalilov2018}.
The convection cell is made of a stainless steel pipe with a $3.5$~mm thick wall.
The inner length of the convection cell is $L=216$~mm and the inner diameter $D=212$~mm.
Both end faces of the convection cell are separated from the heat exchanger chambers by $1$~mm thick copper discs, see a sketch in figure~\ref{PIC1}.
The convection cell is filled with liquid sodium.
The heat exchanger chambers are filled also with liquid sodium and the temperature there is kept constant.
The thin end-face copper plates are intensively washed with liquid sodium from the chamber sides, ensuring homogeneous temperature distributions at their surfaces \citep{Kolesnichenko2017}.
The entire setup is placed on a swing frame, so that the convection cell can be tilted from a vertical position to a horizontal one.
Inclination of the convection cell is then characterised by the angle $ \beta$ between the vertical and the cylinder axis, see figure~\ref{PIC1}.

\begin{table}
\begin{center}
\def~{\hphantom{0}}
\small
\begin{tabular}{ccccc}
&\qquad \qquad& $\lambda$ &\qquad \qquad& $\kappa \times 10^6$ \\
& \qquad& (W/mK) & \qquad& (m$^2$/s) \\
\hline
Stainless steel&& 17  && 4.53 \\
Na && 84.6 && 66.5 \\
Cu  && 391.5 && 111 \\
\end{tabular}
\caption{The thermal conductivity $\lambda$ and the thermal diffusivity $\kappa$ of stainless steel,
liquid sodium (Na) and copper (Cu) at the mean temperature of the experiment of about 410\,K.
}
\label{TAB1}
\end{center}
\end{table}

Obviously, the boundary conditions in a real liquid-metal experiment and the idealised boundary conditions that are considered in numerical simulations, are different.
For example, in the simulations, the cylindrical side wall is assumed to be adiabatic, while in the real experiment it is made from $3.5$~mm thick stainless steel
and is additionally covered by a $30$~mm thick layer of mineral wool.
In table~\ref{TAB1}, the thermal characteristics of stainless steel, liquid sodium and copper are presented.
One can see that the thermal diffusivity of the  stainless steel, although being smaller compared to that of liquid sodium, is not negligible.
Copper, which is known to be the best material for the heat exchangers in the experiments with moderate or high $\Pran$ fluids,
has the thermal diffusivity of the same order as the thermal diffusivity of sodium.
Therefore, massive copper plates would not provide a uniform temperature at the surfaces of the plates \citep{Kolesnichenko2015}.
To avoid this undesirable inhomogeneity, in our experiment, instead of thick copper plates, we use rather thin ones,
which are intensively washed from the outside by liquid sodium of prescribed temperature (see figure~\ref{PIC1}).
The latter process takes place in two heat exchanger chambers, a hot one and a cold one, which are equipped with induction coils.
A sodium flow in each heat exchanger chamber is provided by a travelling magnetic field.
The induction coils are attached near the outer end faces of the corresponding heat exchanger, thus ensuring
that the electromagnetic influence of the inductors on the liquid metal in the convection cell is negligible \citep{Kolesnichenko2017}.
Typical velocities of the sodium flows in the heat exchangers are about $1$~m/s, 
being an order of magnitude higher than the convective velocity inside the convection cell.

In all conducted experiments, the mean temperature of liquid sodium inside the convection cell is about $T_\text{m}=139.8^\circ$C, for which the Prandtl number equals $\Pran \approx0.0093$.
Each experiment is performed for a prescribed and known applied temperature difference $\Theta =T_\text{hot}-T_\text{cold}$,
where $T_\text{hot}$ and $T_\text{cold}$ are the time-averaged temperatures of sodium in, respectively, the hot and cold heat exchanger chambers, which
measured close to the copper plates (see figure~\ref{PIC1}).

In any hot liquid-metal experiment, there exist unavoidable heat losses due to the high temperature of the setup.
To estimate the corresponding power losses $Q_\text{loss}$,
one needs to measure additionally the power, which is required to maintain the same mean temperature $T_\text{m}$ of liquid sodium
inside the convection cell, but under the condition of equal temperatures in both heat exchanger chambers, that is, for $\Theta =0$.
From $Q_\text{loss}$ and the total power consumption $Q$ in the convective experiment by $\Theta\neq0$,
one calculates the effective power $Q_\text{eff}=Q-Q_\text{loss}$ in any particular experiment.

Although the thermal resistance of the two thin copper plates themselves is negligible,
the  sodium-copper  interfaces  provide  additional  effective thermal
resistance of the plates mainly due to inevitable oxide films.
The temperature drop  $\Delta_\text{pl}$ through both copper plates covered by the oxide films
could be calculated then from $ \Delta_\text{pl}=Q_\text{eff}R_\text{pl}$,
as soon as the effective thermal resistance of the plates, $R_\text{pl}$, is known.
Note that for a fixed mean temperature $T_\text{m}$, the value of $R_\text{pl}$ depends on $\Theta$,
since the effective thermal conductivities of the two plates are different due to their different temperatures.

The values of $R_\text{pl}(\Theta)$ for different $\Theta$ are calculated in series
of auxiliary measurements for the case of $\beta=0$ and stable temperature stratification, where the heating is applied from above, to suppress convection.
In this purely conductive case, the effective thermal resistance of the plates, $R_\text{pl}$,
can be calculated from $\Theta$ and the measured effective power $\bar{Q}_\text{eff}$
from the relation $\Theta=\bar{Q}_\text{eff}(R_\text{Na}+R_\text{pl})$,
where the thermal resistance of the liquid sodium equals $R_\text{Na}=L/(\lambda_\text{Na} S)$
with $\lambda_\text{Na}$ being the liquid sodium thermal conductivity and
$S=\pi R^2$ with the cylinder radius $R$.

Using the above measured effective thermal resistances of the plates, $R_\text{pl}(\Theta)$,
in any convection experiment for a given $\Ra$ and $\beta$ and the mean temperature $T_\text{m}$,
the previously unknown temperature drop $\Delta$ inside the convection cell can be calculated
from $\Theta$, $R_\text{pl}(\Theta)$ and measured $Q_\text{eff}$ as follows:
\begin{equation*}
\Delta = \Theta - \Delta_\text{pl}, \quad \Delta_\text{pl}=Q_\text{eff}R_\text{pl}.
\end{equation*}
Note that in all our measurements in liquid sodium, for all considered inclination angles of the convection cell,
the obtained mean temperature and the temperature drop within the cell equal, respectively, $T_\text{m}\approx139.8^\circ$ and $\Delta\approx25.3$\,K.

The Nusselt number is then calculated as
 \begin{equation}
 \Nu = {\frac{L\,Q_\text{eff}}{\lambda_\text{Na} \,S\,\Delta}}.
 \label{nusselt}
 \end{equation}
For a comparison of the experimental results with the DNS and LES results,
where the temperatures at the plates inside the convection cell are known {\it a priori},
the experimental temperature at the hot plate, $T_+$,  and that at the cooled plate, $T_-$, are calculated as follows:
 \begin{equation*}
 T_+=T_\text{m}+\Delta/2, \quad
 T_-=T_\text{m}-\Delta/2.
 \end{equation*}
The Rayleigh number in the experiment is evaluated as 
\begin{eqnarray}
\label{Raexp}
\Ra\equiv \alpha g \Delta D^4/(L\kappa \nu),
\end{eqnarray}
which slightly differs from the value defined in (\ref{Ra}), since $D$ is slightly smaller than $L$.
Thus, for $\alpha=2.56\times10^{-4}$\,K$^{-1}$, $g=9.81$\,m/s$^{2}$, $\Delta=25.3$\,K, $\nu=6.174\times10^{-7}$\,m$^{2}$/s and $\kappa=6.651\times10^{-5}$\,m$^{2}$/s,
the Rayleigh number equals $\Ra=1.42\times10^{7}$.

 \begin{figure}
\centering
\includegraphics{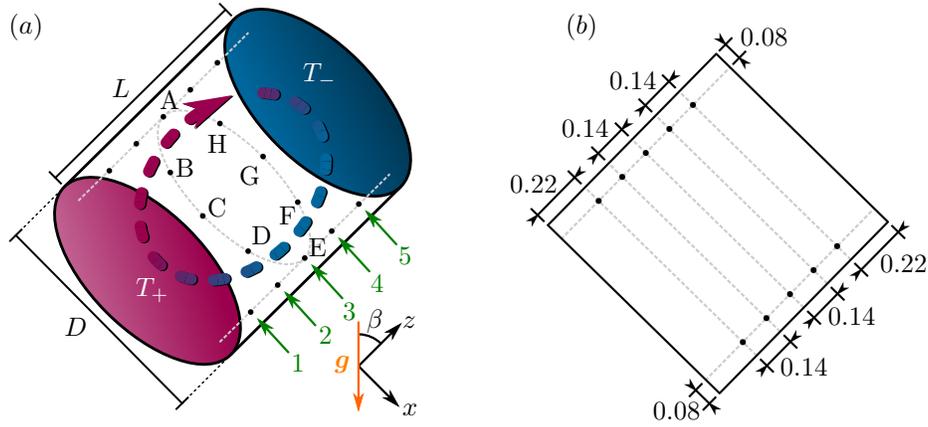}
\caption{
$(a)$ Sketch of an inclined convection cell. $D$ is the diameter and $L$ the height of the cylindrical sample, $\beta$ is the inclination angle,
$T_+$ ($T_-$) the temperature of the heated (cooled) surfaces.
Positioning and naming of the 40 probes inside the cylinder, as considered in the DNS (all combinations of the azimuthal locations A, B, C, D, E, F, G, H and circles $1,\dots,5$)
and 28 probes in the experiments (all combinations of the eight locations A, ..., H for the circles 1, 3 and 5 plus four additional probes: A2, A4, E2, E4).
Note that the azimuthal locations are shown only for the circle 3, not to overload the sketch.
For any inclination angle $\beta>0^\circ$, the upper azimuthal location is A.
$(b)$ Sketch of a central vertical cross-section of the setup from figure $(a)$, with shown distances (normalised with $L=D$)
between the neighbouring probes and between the probes and the side wall of the cylindrical convection cell.
}
\label{PIC2}
\end{figure}

For a deep analysis of the convective liquid sodium flows, the convection cell is equipped with 28 thermocouples, each with an isolated junction of $1$~mm.
The thermocouples are located on 8 lines aligned parallel to the cylinder axis, see figure~\ref{PIC2}.
The azimuthal locations of these lines are distributed with an equal azimuthal step of $45^\circ$
and are marked in figure~\ref{PIC2} by capital letters A to H (counterclockwise, if looking from the cold end face).
The line A has the upper position if $\beta>0^\circ$.
On each of the 8 lines (A to H), 3 or 5 thermocouples are placed.
Thus, all thermocouples are located in five cross-sections of the convection cell, which are parallel to the end faces.
The thermocouples are installed inside the convection cell at the same distance of 17~mm from the inner cylinder side wall
and thus are located on 5 circles, which are marked in figure~\ref{PIC2} with numbers $1,\dots,5$.
The circles 1, 3, and 5 include eight thermocouples, and the circles 2 and 4 only two thermocouples (A and E).

In this paper, besides the measurements by \citet{Khalilov2018}, where a single Rayleigh number $\Ra=(1.42\pm0.03) \times 10^7$ was considered,
we present and analyse also new experimental data, which are
obtained for a certain range of the Rayleigh number, based on different imposed temperature gradients.

\subsection{Direct numerical simulations}

The problem of inclined thermal convection within the Oberbeck--Boussinesq (OB) approximation, which is studied in the DNS,
is defined by the following Navier--Stokes, temperature and continuity equations in cylindrical coordinates $(r, \phi, z)$:
\begin{eqnarray}
D_t \vec{u} &=&  \nu\vec{\nabla}^2 \vec{u}  -\vec{\nabla} p + \alpha g (T-T_0) \hat{\vec{e}}, \label{eq:NS1}\\
D_t T &=&  \kappa \vec{\nabla}^2 T,\label{eq:NS2}\\
\vec{\nabla} \cdot \vec{u} &=& 0, \label{eq:NS3}
\end{eqnarray}
where $D_t$ denotes the substantial derivative,
$\vec{u} = (u_r, u_\phi, u_z)$ the velocity vector field,
with the component $u_z$ in the direction $z$, which is orthogonal to the plates,
$p$ is the reduced kinetic pressure,
$T$ the temperature,
$T_0=(T_++T_-)/2$
and $\hat{\vec{e}}$ is the unit vector, $\hat{\vec{e}}=(-\sin(\beta)\cos(\phi), \sin(\beta)\sin(\phi), \cos(\beta))$.
Within the considered OB approximation, it is assumed that the fluid properties are independent of the temperature and pressure,
apart from the buoyancy term in the Navier--Stokes equation, where the density is taken linearly dependent on the temperature.

These equations are non-dimensionalized by using the cylinder radius $R$,
the free-fall velocity $U_f$, the free-fall time $t_f$,
\begin{eqnarray*}
U_f\equiv(\alpha gR\Delta)^{1/2},\qquad
t_f=R(\alpha gR\Delta)^{-1/2},
\end{eqnarray*}
and the temperature drop between the heated plate and the cooled plate, $\Delta$, as units of length, velocity, time and temperature, respectively.

\begin{table}
    \begin{center}
        \def~{\hphantom{0}}
        \small
        \begin{tabular}{lccrrrrrccr}
            & $\qquad\Ra\qquad$ & $\quad\Pran\quad$ & $\quad\beta\;$ & $\quad t_{\text{avg}}/t_f$ & $\ N_r\ $ & $\ N_\phi\ $ & $\ N_z\ $& $\ \mathcal{N}_{\text{th}}\ $&$\
            \mathcal{N}_{\text{v}}\ $ & $\delta_{\text{v}}/L\qquad$\\ \hline
            DNS\quad& $1.67\times 10^7$ & $0.0094$& $0^\circ$  &  106.9 & 384 & 512 & 768 & 82 & 7 & $3.4\times10^{-3}$ \\
            &&                                    & $36^\circ$ &   53.1 & 384 & 512 & 768 & 67 & 7 & $3.2\times10^{-3}$ \\
            &&                                    & $72^\circ$ &   50.4 & 384 & 512 & 768 & 69 & 7 & $3.2\times10^{-3}$ \\
            &&                                    & $90^\circ$ &   51.4 & 384 & 512 & 768 & 77 & 7 & $3.4\times10^{-3}$ \\\hline
             LES & $1.5 \times 10^7$    & 0.0093   & $0^\circ$  &  460.5 & 100 & 160 & 200 & 32 & 3 & $3.3\times10^{-3}$\\
            &&                                    & $10^\circ$ &  629.4 & 100 & 160 & 200 & 31 & 3 & $3.2\times10^{-3}$\\
            &&                                    & $20^\circ$ &  806.0 & 100 & 160 & 200 & 29 & 3 & $3.1\times10^{-3}$\\
            &&                                    & $30^\circ$ &  468.2 & 100 & 160 & 200 & 27 & 3 & $3.1\times10^{-3}$\\
            &&                                    & $40^\circ$ &  468.2 & 100 & 160 & 200 & 27 & 3 & $3.1\times10^{-3}$\\
            &&                                    & $50^\circ$ &  928.8 & 100 & 160 & 200 & 26 & 3 & $3.1\times10^{-3}$\\
            &&                                    & $60^\circ$ &  560.3 & 100 & 160 & 200 & 27 & 3 & $3.1\times10^{-3}$\\
            &&                                    & $70^\circ$ &  460.5 & 100 & 160 & 200 & 27 & 3 & $3.1\times10^{-3}$\\
            &&                                    & $80^\circ$ &  537.3 & 100 & 160 & 200 & 28 & 3 & $3.2\times10^{-3}$\\
            &&                                    & $90^\circ$ &  652.4 & 100 & 160 & 200 & 30 & 3 & $3.3\times10^{-3}$\\ \hline
            DNS & $1.67\times 10^6$ & $0.0940$    & $0^\circ$  & 4000   &  95 & 128 & 192 & 13 & 3 & $1.21\times10^{-2}$\\
            &&                                    & $9^\circ$  & 1000   &  95 & 128 & 192 & 12 & 3 & $1.24\times10^{-2}$\\
            &&                                    & $18^\circ$ & 2000   &  95 & 128 & 192 & 12 & 3 & $1.19\times10^{-2}$\\
            &&                                    & $27^\circ$ & 1000   &  95 & 128 & 192 & 11 & 3 & $1.15\times10^{-2}$\\
            &&                                    & $36^\circ$ & 2000   &  95 & 128 & 192 & 11 & 3 & $1.13\times10^{-2}$\\
            &&                                    & $54^\circ$ & 2000   &  95 & 128 & 192 & 11 & 3 & $1.15\times10^{-2}$\\
            &&                                    & $72^\circ$ & 1000   &  95 & 128 & 192 & 11 & 3 & $1.11\times10^{-2}$\\
            &&                                    & $90^\circ$ & 1000   &  95 & 128 & 192 & 13 & 3 & $1.11\times10^{-2}$\\ \hline
             DNS & $10^9$            & 1          & $0^\circ$  &  420   & 384 & 512 & 768 & 16 & 11 & $5.2\times10^{-3}$\\
            &&                                    & $9^\circ$  &  385   & 384 & 512 & 768 & 15 & 10 & $5.0\times10^{-3}$\\
            &&                                    & $18^\circ$ &  403   & 384 & 512 & 768 & 15 & 10 & $4.6\times10^{-3}$\\
            &&                                    & $27^\circ$ &  465   & 384 & 512 & 768 & 15 &  9 & $4.3\times10^{-3}$\\
            &&                                    & $36^\circ$ &  447   & 384 & 512 & 768 & 15 &  8 & $3.9\times10^{-3}$\\
            &&                                    & $45^\circ$ &  356   & 384 & 512 & 768 & 15 &  7 & $3.6\times10^{-3}$\\            
            &&                                    & $54^\circ$ &  366   & 384 & 512 & 768 & 15 &  7 & $3.2\times10^{-3}$\\
            &&                                    & $63^\circ$ &  382   & 384 & 512 & 768 & 15 &  6 & $2.9\times10^{-3}$\\            
            &&                                    & $72^\circ$ &  448   & 384 & 512 & 768 & 16 &  6 & $2.8\times10^{-3}$\\
            &&                                    & $81^\circ$ &  182   & 384 & 512 & 768 & 15 &  6 & $2.7\times10^{-3}$\\            
            &&                                    & $90^\circ$ &   36   & 384 & 512 & 768 & 17 &  6 & $2.6\times10^{-3}$
        \end{tabular}
        \caption{Details on the conducted DNS and LES, including the
            time of statistical averaging, $t_{\text{avg}}$, normalised with
            the free-fall time $t_f$; number of nodes $N_r$, $N_\phi$,
            $N_z$ in the directions $r$, $\phi$ and $z$, respectively;
            the number of the nodes within the thermal boundary layer,
            $\mathcal{N}_{\text{th}}$, and within the viscous boundary layer,
            $\mathcal{N}_{\text{v}}$, and the relative thickness of the viscous boundary layer, $\delta_{\text{v}}/L$.
        }
        \label{TAB2}
    \end{center}
\end{table}

To close the system (\ref{eq:NS1})--(\ref{eq:NS3}), the following boundary conditions are considered: no-slip for the velocity at all boundaries, $\vec{u}=0$,
constant temperatures ($T_-$ or $T_+$) at the face ends of the cylinder and adiabatic boundary condition at the side wall, $\partial T/\partial r=0$.

The resulting dimensionless equations are solved numerically with the finite-volume computational code {\sc goldfish},
which uses high-order interpolation schemes in space and a direct solver for the pressure \citep{Kooij2018}.
No turbulence model is applied in the simulations.
The utilised staggered computational grids of about $1.5\times10^8$ nodes, which are clustered near all rigid walls,
are sufficiently fine to resolve the Kolmogorov microscales, see table~\ref{TAB2}.

Since the simulations on such fine meshes are extremely expensive, only 4 inclination angles are considered
for the main case of $\Pran=0.0094$ and $\Ra=1.67\times10^7$, which are $\beta=0^\circ$, $\beta=36^\circ$,  $\beta=72^\circ$ and $\beta=90^\circ$.
To study similarities of the flows with respect to the global heat transport and global flow structures
for a fixed Grashof number, $\Gr\equiv\Ra/\Pran$, and for a fixed value of $\Ra\,\Pran$,
some additional DNS of IC were conducted for the combinations of $\Pran=0.094$ with $\Ra=1.67\times10^6$ and $\Pran=1$ with $\Ra=10^9$. 
In the former additional DNS, 8 different inclination angles are considered,
while in  the latter additional DNS, 11 different values of $\beta$ are examined, see table~\ref{TAB2}.

\subsection{Large Eddy Simulations}

At any fixed time slice, LES generally require more computational efforts per computational node, than the DNS.
However, since the LES are relieved from the requirement to resolve the spatial Kolmogorov microscales,
one can use significantly coarser meshes in the LES compared to those in the DNS,
as soon as the LES are verified against the measurements from the physical point of view
and against the DNS from the numerical point of view.
Thus, the verified LES open a possibility to obtain reliable data faster, compared to the DNS, using modest computational resources.

In our study, the OB equations (\ref{eq:NS1})--(\ref{eq:NS3})
of thermogravitational convection with the LES approach for small-scale turbulence modelling are solved numerically
using the open-source package OpenFOAM~4.1~\citep{Weller1998}
for $\Pran=0.0093$ and $\Ra=1.5\times10^7$
and 10 different inclination angles, equidistantly distributed between $\beta=0^\circ$ and $\beta=90^\circ$.

The package is configured as follows.
The used LES model is that by Smagorinsky-Lilly~\citep{Deardorff1970} with the Smagorinsky constant $C_s = 0.17$.
The turbulent Prandtl number in the core part of the domain equals $\Pran_t = 0.9$ and smoothly vanishes close to the rigid walls.
The utilised finite-volume solver is {\it buoyantBoussinesqPimpleFoam} with PISO algorithm by \citet{Issa1986}.
Time integration is realised with the implicit Euler scheme; linearly are treated also the diffusive and convective terms
(more precisely, using the {\it filteredLinear} scheme).
The resulting systems of linear equations are solved with the Preconditioned Conjugate Gradient (PCG) method 
with the Diagonal Incomplete-Cholesky (DIC) preconditioner for the pressure and 
Preconditioned Biconjugate Gradient (PBiCG) method with 
the Diagonal Incomplete-LU (DILU) preconditioner for other flow components \citep{Fletcher1976, Ferziger2002}. 

All simulations are carried out on a collocated non-equidistant computational grid consisting of 2.9 million nodes, see table~\ref{TAB2}.
The grid has a higher density of the nodes near the boundaries, in order to resolve the boundary layers.
Further details on the numerical method, construction of the computational grid and model verification can be found in \cite{Mandrykin2018}.

\subsection{Methods of data analysis}

\subsubsection{Nusselt number and Reynolds number}

The main response characteristics of the convective system are the global heat and momentum transport represented by the dimensionless Nusselt number $\Nu$
and Reynolds number $\Rey$, respectively.
Within the OB approximation, the Nusselt number equals
\begin{eqnarray}
\Nu=\avg{\Omega_z}{z},
\label{NuOB}
\end{eqnarray}
where $\Omega_z$ is a component of the heat flux vector along the cylinder axis,
\begin{eqnarray}
\Omega_z\equiv \frac{{u_zT}-\kappa\partial_z{T}}{\kappa\Delta/L},
\label{Omega}
\end{eqnarray}
and $\langle \cdot\rangle_{z}$ denotes the average in time and over a cross-section at any distance $z$ from the heated plate.

The Reynolds number can be defined in different ways and one of the common definitions is based on the total kinetic energy of the system:
\begin{equation}
\Rey \equiv (L/\nu)
\sqrt{\avg{{\vec u}\cdot{\vec u}}{}}.
\label{Reynolds-kinetic-energy}
\end{equation}
Here $\vtavg{\cdot}$ denotes the averaging in time and over the entire volume.
We consider also the large-scale Reynolds number
\begin{equation}
\Rey_{\rm U} \equiv {(L/\nu)\sqrt{\avg{\avg{\vec{u}}{t}^2}{V}}}{},
\label{eq:Re-lsc}
\end{equation}
where $\avg{\cdot}{t}$ denotes the averaging in time and
$\avg{\cdot}{V}$ the averaging over the whole convection cell.
Following \citet{Teimurazov2017}, we also evaluate the Reynolds number based on the volume-averaged velocity fluctuations, or small-scale Reynolds number, as
\begin{equation}
\Rey_{u^\prime} \equiv  {(L/\nu)\sqrt{\avg{(\vec{u}-\avg{\vec{u}}{t})^2}{}}}{}.
\label{eq:Re-rms}
\end{equation}
Finally, one can calculate the Reynolds number based on the "wind of turbulence" as follows:
\begin{equation}
\Rey_{\rm w}=(L/\nu)\max_z U_{\rm w}(z),
\label{eq:Re-wind}
\end{equation}
where the velocity of the wind, which is parallel to the heated or cooled plates, can be estimated by
\begin{equation}
U_{\rm w}(z)=\sqrt{\langle u_\phi^2+u_r^2\rangle_{z}},
\label{eq:wind}
\end{equation}
where $u_\phi$ and $u_r$ are the azimuthal and radial components of the velocity, respectively.

In the experiments, the Reynolds number is evaluated based on the average of the estimated axial velocities between the 
probes along the positions A and E.
The velocities are estimated as it is written below, in section 2.4.4.

\subsubsection{Boundary-layer thicknesses}

Close to the heated and cooled plates, the thermal and viscous boundary layers develop.
The thickness of the thermal boundary layer is calculated as
\begin{equation}
\delta_\theta=L/(2\Nu).
\label{eq:deltath}
\end{equation}

Using the slope method, from the $U_{\rm w}(z)$ profile along the cylinder axis, see equation (\ref{eq:wind}),
the viscous boundary-layer thickness $\delta_u$ is calculated as follows:
\begin{equation}
\delta_u=\max_z \{U_{\rm w}(z)\}\left(\left.\frac{dU_{\rm w}}{dz}\right\vert_{z=0}\right)^{-1}.
\end{equation}

\subsubsection{Properties of the large-scale circulation}

In the DNS and LES, the information on all flow components is available at every small finite volume associated with any grid node.
In the experiments, all the information about the flow structures is obtained from the 28 temperature probes located as shown in figure~\ref{PIC2} and discussed in section 2.1.
The probes are placed in five different horizontal cross-sections of the cylindrical sample, which are parallel to the heated or cooled surfaces.
To make a comparison between the DNS, LES and the experiment possible, 
we measure the temperature and monitor its temporal evolution at the same locations in all three approaches.
The only difference is that in the DNS and LES there are 8 virtual probes in each cross-section, 
while in the experiment, there are 8 probes in the cross-sections 1, 3 and 5 and only 2 probes in the cross-sections 2 and 4.
The azimuthal locations A to H in the experiment, DNS and LES are exactly the same.

From the temperature measurements at the above discussed locations,
one can evaluate the phase and the strength of the so-called wind of turbulence, or large scale circulation (LSC).
To do so, the method by \cite{Cioni1997} is applied, which is widely used in the RBC experiments \citep{Brown2006, Xi2007, He2016, Bai2016, Khalilov2018} and simulations
\citep{Mishra2011, Stevens2011, Wagner2012, Ching2017}.
Thus, the temperature measured at 8 locations in the central cross-section, from A3 to H3 along the central circle 3, is fitted by the cosine function 
\begin{equation}
\label{theta}
T(\theta) = T_{\rm m} + \delta_3\cos(\theta-\theta_3),
\end{equation}
to obtain the orientation of the LSC, represented by the phase $\theta_3$, 
and the strength of the first temperature mode, i.e., the amplitude $\delta_3$, which indicates the temperature drop between the opposite sides of the cylinder side wall.
At the warmer part of the side wall, the LSC carries the warm plumes from the heated plate towards the cold plate
and on the opposite colder part, it carries the cold plumes in the opposite direction.

In a similar way one can evaluate the LSC phases $\theta_1$ and $\theta_5$ and the strengths $\delta_1$ and $\delta_5$ at other heights from the heated plates,
i.e., along the circle 1 (closer to the heated plate) and along the circle 5 (closer to the cooled plate), respectively.

\subsubsection{Velocity estimates}

While in the DNS and LES the spatial distributions of all velocity components are available, 
the direct measurements of the velocity in the experiments on natural thermal convection in liquid sodium remain to be impossible so far.
In order to estimate the velocities from the temperature measurements in the experiment, 
the cross-correlations for all combinations of any two neighbouring probes along the azimuthal locations from A to H are used. 

For example, the normalised cross-correlation function $\left.C\right|_{A1,\,A2}(\tau)$
for the temporal dependences of the temperatures $\left.T\right|_{A1}$ and $\left.T\right|_{A2}$ measured by the probes at the locations A1 and A2 is calculated as follows:
\begin{equation}
\left.C\right|_{A1,\,A2}(\tau)
\propto
 \sum_{j} (\left.T\right|_{A1}(t_j)-\langle \left.T\right|_{A1}\rangle_t )\cdot (\left.T\right|_{A2}(t_j+\tau)-\langle \left.T\right|_{A2}\rangle_t).
\end{equation}
The first maximum of the function $\left.C\right|_{A1,\,A2}(\tau)$ at $\tau=\tau_c$ provides the correlation time $\tau_c$.
From the known distance between the probes A1 and A2 and the estimated time, $\tau_c$, which is needed for the flow to bring a thermal plume from the location A1 to A2,
one can estimate the mean velocity of the flow between the locations A1 and A2.

In a similar way one estimates the mean velocities between the probes A2 and A3, etc., along the azimuthal location A.
The  mean velocities along the other azimuthal locations, from B to H, are calculated analogously.

\section{Results and discussion}

\begin{table}
    \begin{center}
        \def~{\hphantom{0}}
        \small
        \begin{tabular}{lccrrc|cccrr}
& $\quad\Ra\quad$ & $\quad\Pran\quad$ & $\quad\beta\;$ & $\quad \Nu\quad$ &$\quad$&$\quad$&$\quad\Ra\qquad$ & $\quad\Pran\quad$ & $\quad\beta\;$ & $\quad \Nu\quad$ \\ \hline
    Experiment&  $5.27\times 10^6$ & $0.0094$& $0^\circ$ & 4.97 &&& $1.42\times 10^7$ & $0.0093$ & $20^\circ$ & 6.51 \\ 
&  $6.43\times 10^6$ & $0.0097$& $0^\circ$ & 5.18 &&& $1.42\times 10^7$ & $0.0093$ & $30^\circ$ & 7.02 \\ 
&  $9.32\times 10^6$ & $0.0096$& $0^\circ$ & 5.53 &&& $1.42\times 10^7$ & $0.0093$ & $40^\circ$ & 7.07 \\ 
&  $1.12\times 10^7$ & $0.0095$& $0^\circ$ & 5.75 &&& $1.42\times 10^7$ & $0.0093$ & $50^\circ$ & 7.15 \\ 
&  $1.18\times 10^7$ & $0.0093$& $0^\circ$ & 5.84 &&& $1.42\times 10^7$ & $0.0093$ & $60^\circ$ & 7.23 \\ 
&  $1.28\times 10^7$ & $0.0094$& $0^\circ$ & 5.91 &&& $1.42\times 10^7$ & $0.0093$ & $70^\circ$ & 7.30 \\ 
&  $1.42\times 10^7$ & $0.0093$& $0^\circ$ & 6.04 &&& $1.42\times 10^7$ & $0.0093$ & $80^\circ$ & 7.13 \\ 
&  $1.43\times 10^7$ & $0.0091$& $0^\circ$ & 6.09 &&& $6.52\times 10^6$ & $0.0095$ & $90^\circ$ & 5.87 \\ 
&  $1.55\times 10^7$ & $0.0093$& $0^\circ$ & 6.18 &&& $8.91\times 10^6$ & $0.0094$ & $90^\circ$ & 6.19 \\
&  $1.80\times 10^7$ & $0.0091$& $0^\circ$ & 6.39 &&& $1.11\times 10^7$ & $0.0093$ & $90^\circ$ & 6.53 \\
&  $2.06\times 10^7$ & $0.0088$& $0^\circ$ & 6.79 &&& $1.32\times 10^7$ & $0.0091$ & $90^\circ$ & 6.77 \\
&  $2.18\times 10^7$ & $0.0088$& $0^\circ$ & 6.55 &&& $1.42\times 10^7$ & $0.0093$ & $90^\circ$ & 6.84 \\
&  $2.37\times 10^7$ & $0.0086$& $0^\circ$ & 6.92 &&& $1.60\times 10^7$ & $0.0090$ & $90^\circ$ & 7.07 \\
&  $1.42\times 10^7$ & $0.0093$& $10^\circ$& 6.17 &&& $1.88\times 10^7$ & $0.0086$ & $90^\circ$ & 7.47 \\ \hline
         DNS          & $1.67\times 10^7$ & $0.0094$& $0^\circ$ &  9.66 &&& $1.67\times 10^7$ & $0.0094$& $72^\circ$  & 11.91\\
                      & $1.67\times 10^7$ & $0.0094$& $36^\circ$& 12.24 &&& $1.67\times 10^7$ & $0.0094$& $90^\circ$  & 10.38\\\hline
          LES        & $1.5 \times 10^7$  & $0.0093$& $0^\circ$ &  9.27 &&& $1.5 \times 10^7$ & $0.0093$& $50^\circ$ & 11.95  \\
                     & $1.5 \times 10^7$  & $0.0093$& $10^\circ$&  9.65 &&& $1.5 \times 10^7$ & $0.0093$& $60^\circ$ & 11.80  \\
                     & $1.5 \times 10^7$  & $0.0093$& $20^\circ$& 10.28 &&& $1.5 \times 10^7$ & $0.0093$& $70^\circ$ & 11.61  \\
                     & $1.5 \times 10^7$  & $0.0093$& $30^\circ$& 11.41 &&& $1.5 \times 10^7$ & $0.0093$& $80^\circ$ & 10.97  \\
                     & $1.5 \times 10^7$  & $0.0093$& $40^\circ$& 11.71 &&& $1.5 \times 10^7$ & $0.0093$& $90^\circ$ & 10.06  \\ \hline
            DNS      & $1.67\times 10^6$ & $0.0940$ & $0^\circ$  & 8.16 &&& $1.67\times 10^6$ & $0.0940$& $36^\circ$ & 9.79 \\
                     & $1.67\times 10^6$ & $0.0940$ & $9^\circ$  & 8.78 &&& $1.67\times 10^6$ & $0.0940$& $54^\circ$ & 9.95 \\
                     & $1.67\times 10^6$ & $0.0940$ & $18^\circ$ & 9.23 &&& $1.67\times 10^6$ & $0.0940$& $72^\circ$ & 9.55 \\
                     & $1.67\times 10^6$ & $0.0940$ & $27^\circ$ & 9.64 &&& $1.67\times 10^6$ & $0.0940$& $90^\circ$ & 8.55 \\ \hline
            DNS      & $10^9$ & 1 & $ 0^\circ$ & 63.74 &&& $10^9$ & 1 & $54^\circ$ & 67.24 \\
			         & $10^9$ & 1 & $ 9^\circ$ & 64.58 &&& $10^9$ & 1 & $63^\circ$ & 64.78 \\
                     & $10^9$ & 1 & $18^\circ$ & 65.81 &&& $10^9$ & 1 & $72^\circ$ & 62.53 \\
                     & $10^9$ & 1 & $27^\circ$ & 65.58 &&& $10^9$ & 1 & $54^\circ$ & 59.60 \\
                     & $10^9$ & 1 & $36^\circ$ & 66.05 &&& $10^9$ & 1 & $90^\circ$ & 57.52 \\
                     & $10^9$ & 1 & $45^\circ$ & 67.15 &&& & & & 
        \end{tabular}
                \caption{Nusselt numbers, as they were obtained in the experiments, DNS and LES.}
        \label{TAB3}
    \end{center}
\end{table}

In this section, we directly compare the results for inclined convection (IC) in a cylindrical sample of the diameter-to-height aspect ratio one,
as they were obtained in the liquid-sodium DNS for $\Ra=1.67\times 10^7$ and $\Pran=0.0094$,
LES for $\Ra=1.5\times 10^7$ and $\Pran=0.0093$ and liquid-sodium experiments for $\Ra=1.42\times 10^7$ ($\Pran\approx0.0093$),
see tables~\ref{TAB3} and~\ref{TAB4}.

Further liquid-sodium experiments were conducted to measure the scaling relations of the Nusselt number versus the Rayleigh number 
in the RBC (the inclination angle $\beta=0^\circ$) and VC configurations ($\beta=90^\circ$),
for the $\Ra$-range around $\Ra=10^7$.

\begin{table}
    \begin{center}
        \def~{\hphantom{0}}
        \small
        \begin{tabular}{lccrrrr}
            & $\qquad\Ra\qquad$ & $\quad\Pran\quad$ & $\qquad\beta\;$ & $\qquad\Rey\quad$ & $\qquad\Rey_{u^\prime}\;$ & $\qquad\Rey_{\rm U}\;$ \\ \hline
            DNS\qquad \;& $1.67\times 10^7$ & $0.0094$         & $0^\circ$  & 17927 & 12828 & 12523 \\
            &&                                    & $36^\circ$    & 19430 & 10081 & 16609 \\
            &&                                    & $72^\circ$    & 13372 &  5527 & 12176 \\
            &&                                    & $90^\circ$    & 10110 &  4406 &  9100 \\\hline
            LES & $1.5 \times 10^7$    & 0.0093   & $0^\circ$   & 16271 & 11363 & 11646\\ 
            &&                                    & $10^\circ$  & 17015 & 11454 & 12582 \\ 
            &&                                    & $20^\circ$  & 17658 & 11128 & 13711\\ 
            &&                                    & $30^\circ$  & 17722 & 9699  & 14832\\ 
            &&                                    & $40^\circ$  & 17230 & 8241  & 15131\\    
            &&                                    & $50^\circ$  & 16760 & 7106  & 15178\\ 
            &&                                    & $60^\circ$  & 15167 & 6022  & 13920\\ 
            &&                                    & $70^\circ$  & 13114 & 5319  & 11987\\ 
            &&                                    & $80^\circ$  & 11472 & 4598  & 10511\\ 
            &&                                    & $90^\circ$   & 9602  & 3845  & 8798\\ \hline    
            DNS & $1.67\times 10^6$ & $0.0940$    & $0^\circ$    & 1326  & 1241  &  467 \\
            &&                                    & $9^\circ$   & 1421  & 1103  & 1103 \\
            &&                                    & $18^\circ$    & 1460  &  809  & 1216 \\
            &&                                    & $27^\circ$   & 1457  &  702  & 1277 \\
            &&                                    & $36^\circ$   & 1430  &  634  & 1281 \\
            &&                                    & $54^\circ$    & 1328  &  366  & 1277 \\
            &&                                    & $72^\circ$   &  991  &   77  &  988 \\
            &&                                    & $90^\circ$   &  725  &    0  &  725\\\hline 
            DNS & $10^9$ & $1$                    & $ 0^\circ$ & 4721 & 4254 & 2047 \\ 
            &&                                    & $ 9^\circ$ & 5126 & 3624 & 3625 \\
            &&                                    & $18^\circ$ & 4968 & 2504 & 4291 \\
            &&                                    & $27^\circ$ & 4648 & 2144 & 4124 \\
            &&                                    & $36^\circ$ & 4063 & 1905 & 3589 \\            
            &&                                    & $45^\circ$ & 3785 & 1836 & 3310 \\            
            &&                                    & $54^\circ$ & 3307 & 1676 & 2851 \\            
            &&                                    & $63^\circ$ & 2439 & 1288 & 2070 \\
            &&                                    & $72^\circ$ & 1661 &  752 & 1481 \\
            &&                                    & $81^\circ$ & 1226 &  292 & 1191 \\
            &&                                    & $90^\circ$ &  839 &   12 &  839
            
        \end{tabular}
        \caption{Reynolds numbers, as they were obtained in the DNS and LES, see the definitions (\ref{Reynolds-kinetic-energy}), (\ref{eq:Re-lsc}) and (\ref{eq:Re-rms}).
        }
        \label{TAB4}
    \end{center}
\end{table}

Additionally, we make a comparison with the auxiliary DNS results for $\Ra=1.67\times10^6$ and $\Pran=0.094$, 
where the product of the Rayleigh number and Prandtl number, $\Ra\,\Pran\approx 1.57\times10^5$, 
is the same as in the main DNS for liquid sodium with $\Ra=1.67\times 10^7$ and $\Pran=0.0094$, see tables~\ref{TAB3} and~\ref{TAB4}.
This auxiliary case is interesting by the following reasons.
The ratio of the thermal diffusion time scale, $t_\kappa=R^2/\kappa$, to the free-fall time scale, $t_f=\sqrt{R/\alpha g\Delta}$, 
in both cases is the same, since $t_\kappa/t_f\sim \sqrt{\Ra\,\Pran}$.
In contrast, in this auxiliary DNS case, the ratio $t_\nu/t_f$ of the viscous diffusion time scale $t_\nu=R^2/\nu$ to the free-fall time scale $t_f$
is about tenfold smaller than that in the main liquid-sodium DNS, albeit being about tenfold larger than $t_\kappa/t_f$.
Hence, thermal diffusion dominates over viscous diffusion in both considered sets of parameters
and for similar $\Ra\,\Pran$ one might expect similar global temperature distributions and quantitatively similar heat and momentum transport in IC.
Note that in the liquid-sodium DNS, the diffusion times are $t_\kappa \approx 140\,t_f$ and $t_\nu\approx14900\, t_f$.

Another set of auxiliary DNS of IC is conducted for $\Ra=10^9$ and $\Pran=1$ for a comparison.
In this case, the Grashof number, $\Gr\equiv\Ra/\Pran=10^9$, is similar to that in the main liquid-sodium case ($\Gr\approx1.8\times10^9$), 
but for this Prandtl-number-one case and the liquid-sodium case we generally do not expect a close similarity of the global flow characteristics.

A summary of the conducted simulations and experiments can be found in table~\ref{TAB2}.
The free-fall time in the experiments equals $t_f=\sqrt{R/(\alpha g\Delta)}\approx1.3\,$s and is similar to that in the main DNS and LES.
Thus, the conducted DNS cover about two minutes of the real-time experiment only, which was conducted for about 7 hours.
One should note that although the DNS statistical averaging time is quite short, collecting of about $100\,t_f$ statistics for the case $\beta=0^\circ$ 
consumed about $390\,000\,$CPUh at the SuperMUC at the Leibniz Supercomputing Center and required about 60 days of runtime.

In the remaining part of this section we investigate the integral time-averaged quantities like the global heat transport (Nusselt number)
and the global momentum transport (Reynolds number),
provide the evidence of a very good agreement between the simulations and experiment
and present a complementary picture of the dynamics of the large-scale flows in liquid-sodium IC.

\subsection{Time-averaged heat and momentum transport}
\begin{figure}
\centering
\includegraphics{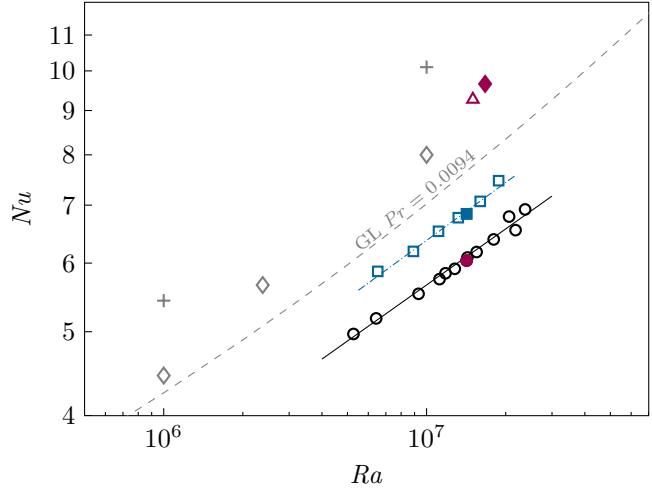}
\caption{Nusselt number versus Rayleigh number, as obtained in
the RBC experiments for different Rayleigh numbers, $\Pran\approx0.009$, $\beta=0^\circ$ (open circles)
with the effective scaling $\Nu\approx 0.177 \Ra^{0.215}$ (solid line);
the RBC experiment for $\Ra=1.42\times 10^7$, $\Pran=0.0093$, $\beta=0^\circ$ (filled circle;
this run is the longest one (7h) in the series of measurements.
For the same $\Ra$ and $\Pran$, the Nusselt numbers were also measured for different $\beta$,
see table~\ref{TAB3}); the VC experiments for different Rayleigh numbers, $\Pran\approx0.009$, $\beta=90^\circ$ (open squares)
with the effective scaling $\Nu\approx 0.178\Ra^{0.222}$ (dash-dotted line);
the DNS for $\Ra=1.67\times 10^7$, $\Pran=0.0094$, $\beta=0^\circ$ (filled diamond);
the LES for $\Ra=1.5\times 10^7$, $\Pran=0.0093$, $\beta=0^\circ$ (open triangle).
Results of the RBC DNS by \cite{Scheel2017} for $\Pr=0.005$ (open diamonds) and for $\Pran=0.025$ (pluses) and
predictions for $\Pran=0.0094$ of the \cite{Grossmann2000, Grossmann2001} theory considered with the pre-factors from \cite{Stevens2013}
(dash line) are presented for comparison.
Everywhere a cylindrical convection cell of the aspect ratio 1 is considered.
}
\label{PIC3}
\end{figure}

First we examine the classical case of RBC without inclination ($\beta=0^\circ$).
The time-averaged mean heat fluxes, represented by the Nusselt numbers, are presented in figure~\ref{PIC3}.

Numerical data, i.e., the DNS and LES for liquid sodium, demonstrate an excellent agreement.
Also in figure \ref{PIC3} we compare our numerical results with the DNS by \citet{Scheel2017},
for $\Pran=0.005$ and $\Pran=0.025$. 
Our numerical results for $\Pran=0.0094$ and $\Pran=0.0093$ take place between the cited results by \citet{Scheel2017}, as expected.
Note that our LES and DNS and the DNS by \citet{Scheel2017} were conducted using completely different codes (Nek5000 spectral element package in the latter case),
which nevertheless lead to very similar results. This verifies the independence of the obtained results from the used numerical method.

For $\Pran=0.0094$, the predictions by \citet{Grossmann2000, Grossmann2001} theory, with the prefactors from \citet{Stevens2013}, 
take place in figure~\ref{PIC3} between the obtained experimental and numerical data.
The experimental data for a certain $\Ra$-range around $\Ra=10^7$, shown in figure~\ref{PIC3}, follow a scaling relation 
\begin{eqnarray}
\label{NuscalingRBC}
\Nu\approx 0.177 \Ra^{0.215}\qquad(\text{RBC,}\quad\beta=0^\circ).
\end{eqnarray}

The experimental data exhibit generally lower Nusselt numbers compared to the numerical data and this can be explained by the following two reasons.
First,  the ideal  boundary  conditions of constant temperatures at the plates
can  not be provided in the experiments, since each emission of a sufficiently strong thermal plume affects, at least for a short  time, 
the local temperature at the plate, which results in a reduction of the averaged heat flux compared to that in the simulations with the  ideal  boundary conditions.
Second, the impossibility to measure the temperature directly at the outer surfaces of the copper plates leads to a slight 
overestimation of $\Theta$ and $\Delta$ (see figure~\ref{PIC2}) and, hence, of the effective Rayleigh numbers for the measured Nusselt numbers.
Anyway, the numerical, theoretical and experimental data for the $\Nu$ versus $\Ra$ scaling are found to follow similar scaling laws.

In figure~\ref{PIC3} we also present the measured scaling relations for $\Nu$ versus $\Ra$ for the case of VC, where the inclination angle equals $\beta=90^\circ$.
The scaling relation in VC is found to be quite similar to that in RBC, namely 
\begin{eqnarray}
\label{NuscalingVC}
\Nu\approx 0.178 \Ra^{0.222}\qquad(\text{VC,}\quad\beta=90^\circ).
\end{eqnarray}
The absolute values of the Nusselt numbers in VC are, however, larger than in RBC.

\begin{figure}
\centering
\includegraphics{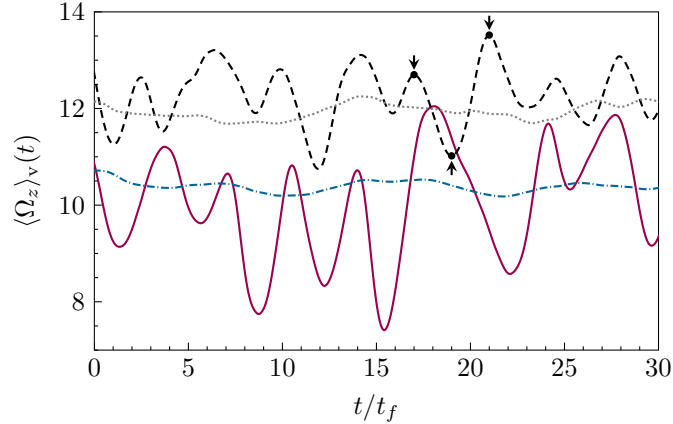}
\caption{Time dependences of the volume-averaged component of the heat flux vector along the cylinder axis,
$\avg{\Omega_z}{V}$, as obtained in the DNS for $\Ra=1.67\times 10^7$,
$\Pran=0.0094$ and four different inclination angles $\beta=0^\circ$ (solid line),
$\beta=36^\circ$ (dash line), $\beta=72^\circ$ (dotted line) and $\beta=90^\circ$ (dash-dotted line).
Time is normalised with $t_f=R(\alpha gR\Delta)^{-1/2}$.
The arrows indicate the dimensionless times, which are marked in figure~\ref{PIC9} with vertical lines and to the snapshots in figure~\ref{PIC10}.}
\label{PIC4}
\end{figure}
\begin{figure}
\centering
\includegraphics{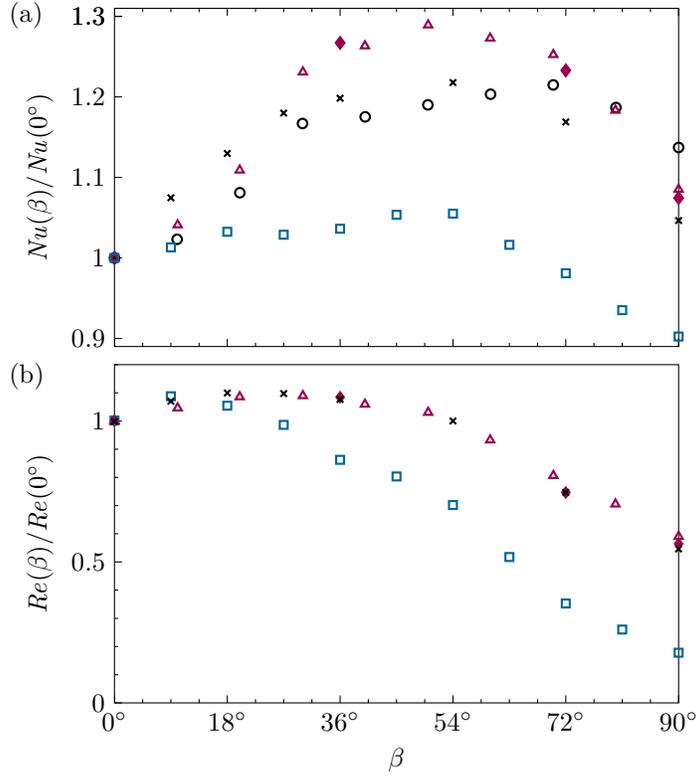}
\caption{$(a)$ Normalised Nusselt number $\Nu(\beta)/\Nu(0^\circ)$ versus the inclination angle $\beta$, as obtained in
the DNS for $\Ra=1.67\times 10^7$, $\Pran=0.0094$ (filled diamonds),
the LES for $\Ra=1.5\times 10^7$, $\Pran=0.0093$ (open triangles),
the DNS for $\Ra=1.67\times 10^6$, $\Pran=0.094$ (crosses),
the experiments for $\Ra=1.42\times 10^7$, $\Pran\approx0.0093$ (open circles) and 
the DNS for $\Ra=10^9$, $\Pran=1$ (squares).
$(b)$ Normalised Reynolds number versus the inclination angle $\beta$, as obtained in
the same DNS, LES and experiments as in $(a)$; similar symbols are used as in $(a)$.}
\label{PIC5}
\end{figure}

In figure~\ref{PIC4}, the time evolution of the volume-averaged components of the heat flux vector along the cylinder axis,
$\avg{\Omega_z}{V}$ are presented, as they are obtained in the DNS for $\Ra=1.67\times 10^7$,
$\Pran=0.0094$ and four different inclination angles between $\beta=0^\circ$ (RBC) and $\beta=90^\circ$ (VC).
Obviously, in the RBC case, the fluctuations of the heat flux around its mean value are extreme
and reach up to $\pm 44\,\%$ of $\avg{\Omega_z}{}$. 
The strength of the fluctuations gradually decreases with growing inclination angle $\beta$ and amount only $\pm 3\,\%$  of $\avg{\Omega_z}{}$ in the VC case.
In figure~\ref{PIC4} one can see that for the inclination angles $\beta=36^\circ$ and $\beta=72^\circ$,
the mean heat transport is stronger than in the RBC or VC cases.
This supports a general tendency that in small-$\Pran$ fluids the heat transport becomes more efficient, when the convection cell is tilted.
Figure~\ref{PIC5}a and table~\ref{TAB3} provide a more detailed evidence of this fact,
based on our measurements and numerical simulations.

In figure~\ref{PIC5}a, the Nusselt numbers in IC are presented, which are normalised by $\Nu$ of the RBC case,
for the same $\Ra$ and $\Pran$, i.e. the dependence of $\Nu(\beta)/\Nu(0^\circ)$ on the inclination angle $\beta$.
Very remarkable is that for similar $\Ra$ and $\Pran$, the DNS and LES deliver very similar values of $\Nu$.
One can see that the numerical data are in good agreement with the experimental data, 
taking into account that the Rayleigh number in the experiment is slightly smaller ($\Ra=1.42\times 10^7$) 
compared to that in the DNS ($\Ra=1.67\times 10^7$). 

As discussed above, we want also to compare our results for liquid sodium with the DNS data for similar $\Ra\,\Pran$ 
and with the DNS data for similar $\Ra/\Pran$, as in our liquid-sodium measurements and numerical simulations. 
Figure~\ref{PIC5}a and table~\ref{TAB3} show that the obtained Nusselt numbers for the same $\Ra\,\Pran$
($\Ra=1.67\times 10^6$, $\Pran=0.094$) are in very good agreement with the liquid-sodium experimental results ($\Ra=1.42\times 10^7$, $\Pran\approx0.0093$).
Remarkable is that not only the relative Nusselt number, $\Nu(\beta)/\Nu(0^\circ)$, but also the absolute values of $\Nu$ are very similar in
the liquid-sodium case and in the case of different $\Pran<1$ but the same $\Ra\,\Pran$. 
In contrast to that, the $\Nu$-dependence on the inclination angle in the case of the same Grashof number is different, as expected.
In that case, the maximal relative increase of the Nusselt number due to the cell inclination is only about 6\%, while in the liquid-sodium case it is up to 29\%.

In figure~\ref{PIC5}b and table~\ref{TAB4}, the results on the Reynolds number are presented for 
the liquid-sodium measurements and simulations as well as for the auxiliary DNS.
Again, the agreement between the experiments, DNS and LES for liquid sodium is excellent.
The dependences  of $\Rey(\beta)/\Rey(0^\circ)$ on the inclination angle, obtained in the liquid-sodium experiments
and in the DNS for the same $\Ra\,\Pran$, demonstrate perfect agreement. 
The values of $\Rey(\beta)/\Rey(0^\circ)$ first slightly increase with the inclination angle and then smoothly decrease, so that 
the Reynolds number $\Rey(90^\circ)$ in the VC case is significantly smaller than the Reynolds number $\Rey(0^\circ)$ in the RBC case.
Here one should notice that the absolute values of $\Rey$ in the liquid-sodium case are significantly larger than
in the IC flows for a similar $\Ra\,\Pran$. 
In the case of the same Grashof number, the Reynolds numbers decrease much faster with growing inclination angle 
than in the liquid-sodium case.
\begin{figure}
\centering
\includegraphics{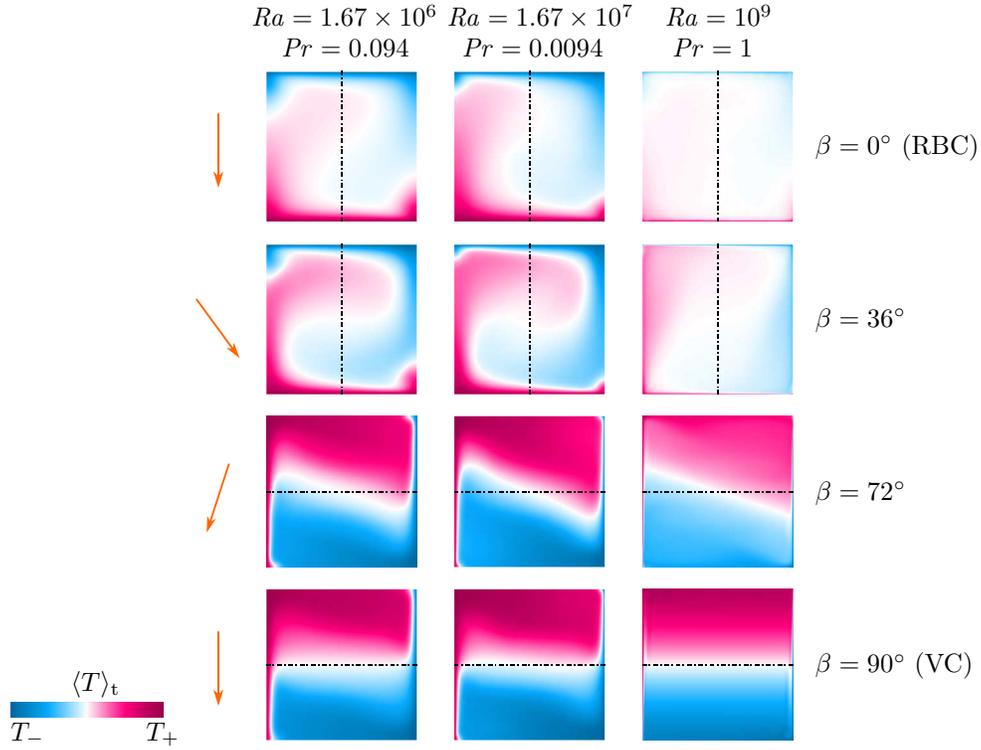}
\caption{Table of vertical slices through the time-averaged temperature in the plane of the LSC,
as obtained in the DNS for $\Ra=1.67\times10^6$, $\Pran=0.094$ (left column);
$\Ra=1.67\times10^7$, $\Pran=0.0094$ (central column) and $\Ra=10^9$, $\Pran=1$ (right column).
From top to bottom, the inclination angle $\beta$ changes from $\beta=0^\circ$ (RBC case)
through $\beta=36^\circ$ and $\beta=72^\circ$ to $\beta=90^\circ$ (VC case).
The black dash-dotted lines indicate the cylinder axis and the arrows on the left show the direction of the gravity vector
in the corresponding row of the temperature slices.}
\label{PIC6}
\end{figure}
\begin{figure}
\centering
\includegraphics{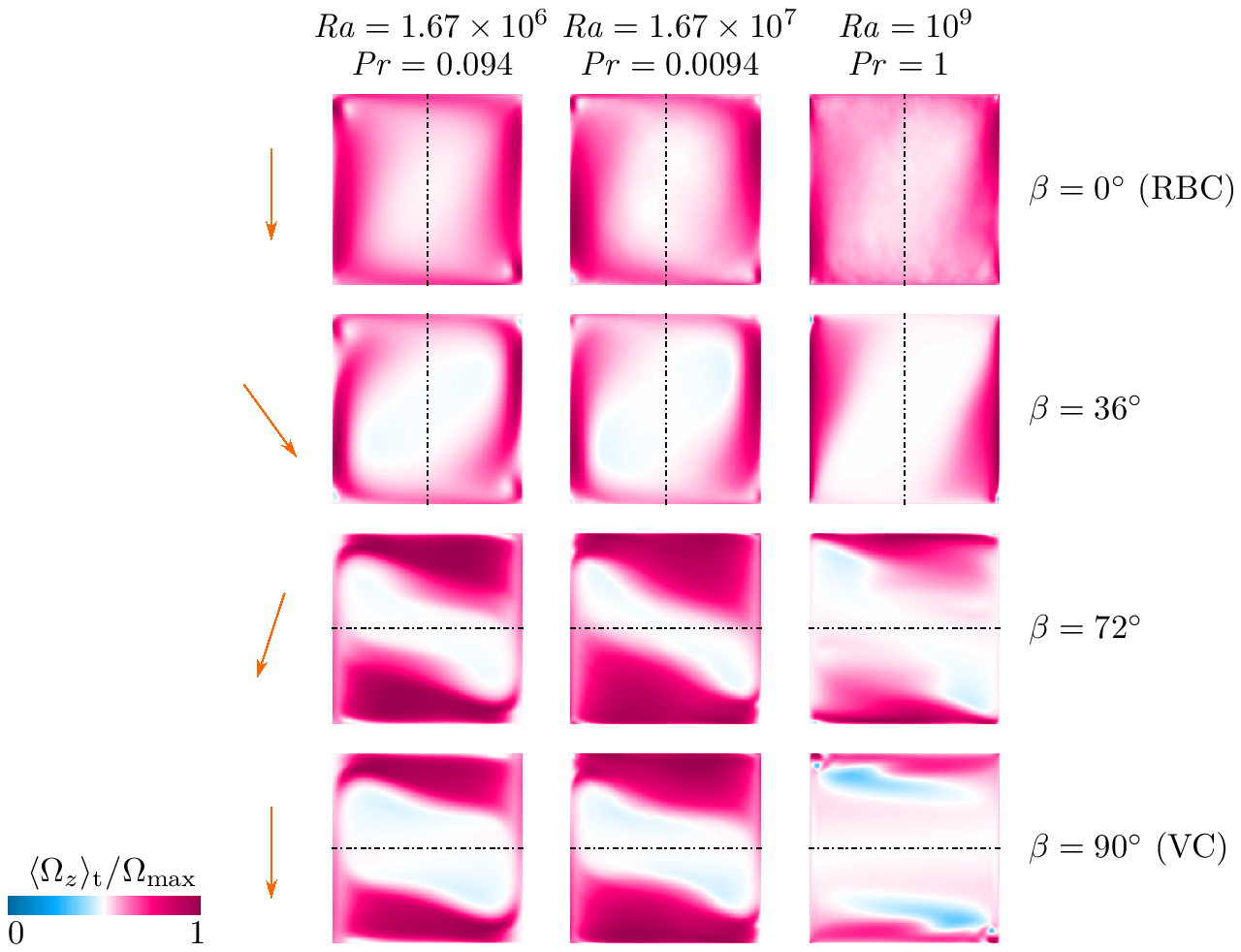}
\caption{Table of vertical slices in the plane of the LSC of the time-averaged
component of the heat flux vector along the cylinder axis,
$\avg{\Omega_z}{t}$, normalised by its maximal value through the entire volume, $\Omega_{\max}$,
as obtained in the DNS for $\Ra=1.67\times10^6$, $\Pran=0.094$ (left column);
$\Ra=1.67\times10^7$, $\Pran=0.0094$ (central column) and $\Ra=10^9$, $\Pran=1$ (right column).
From top to bottom, the inclination angle $\beta$ changes from $\beta=0^\circ$ (RBC case)
through $\beta=36^\circ$ and $\beta=72^\circ$ to $\beta=90^\circ$ (VC case).
The black dash-dotted lines indicate the cylinder axis and the arrows on the left show the direction of the gravity vector
in the corresponding row of the slices.}
\label{PIC7}
\end{figure}

Since the Nusselt numbers and relative Reynolds numbers behave very similar in the liquid-sodium IC experiments and in the 
DNS for similar $\Ra\,\Pran$, we compare the time-averaged flow structures for these cases in figures~\ref{PIC6} and~\ref{PIC7}.
In these figures, the time-averaged temperature (figure~\ref{PIC6}) and the time-averaged component of the heat flux vector along the cylinder axis
$\avg{\Omega_z}{t}$ (figure~\ref{PIC7}) are presented in the plane of the LSC, for different inclination angles.
One can see that both, the temperature distributions and the heat flux distributions, in the liquid-sodium case and in the case of a similar  $\Ra\,\Pran$
look almost identically.
In contrast to them, the corresponding distributions for a similar Grashof number look considerably different.
The difference is especially pronounced for the inclination angle $\beta=36^\circ$.
While in the liquid-sodium flow for $\beta=36^\circ$ there persist two intertwined plumes, a hot one and a cold one, 
the temperature in the Prandtl-number-one case is better mixed (figure~\ref{PIC6}) and 
the heat-flux distribution appears in a form of two triangular-shaped separated spots (figure~\ref{PIC7}).

From the above presented one can conclude that similar Grashof numbers lead neither to similar integral quantities like $\Nu$ or $\Rey$, nor to similar heat flow structures in IC.
In contrast to that, the small-Prandtl-number IC flows of similar $\Ra\,\Pran$
have similar Nusselt numbers $\Nu$, similar relative Reynolds numbers $\Rey(\beta)/\Rey(0^\circ)$
and similar mean temperature and heat flux distributions.

\subsection{Dynamics of the large scale flow}

In this section, we focus on the reconstruction of the rich structural dynamics of the large scale IC flows in liquid sodium.

It is well known from the previous RBC studies that the LSC in RBC can show different azimuthal orientations \citep{Brown2006, Wagner2012}
and can exhibit complicated dynamics with twisting \citep{Funfschilling2004, Funfschilling2008, He2016}  and sloshing \citep{Xi2006, Xi2009, Zhou2009,Brown2009, Bai2016}.
In very-low-Prandtl-number RBC, this complicated behaviour of the LSC was reported in the experiments in mercury by \cite{Cioni1997} and in the simulations by \cite{Schumacher2016, Scheel2016, Scheel2017}.

In our simulations and experiments in liquid sodium, we observe the twisting and sloshing dynamics of the LSC in the RBC configuration of the flow, i.e. without any cell inclination, 
as well as for small inclination angles $\beta$ until a certain critical $\beta=\beta_s$.
The experimental data suggest that a transition to the non-sloshing behaviour of the LSC is quite sharp and 
it is presumably caused by the increasing stratification of the temperature at larger inclination angles \citep{Khalilov2018}.
\begin{figure}
\centering
\includegraphics{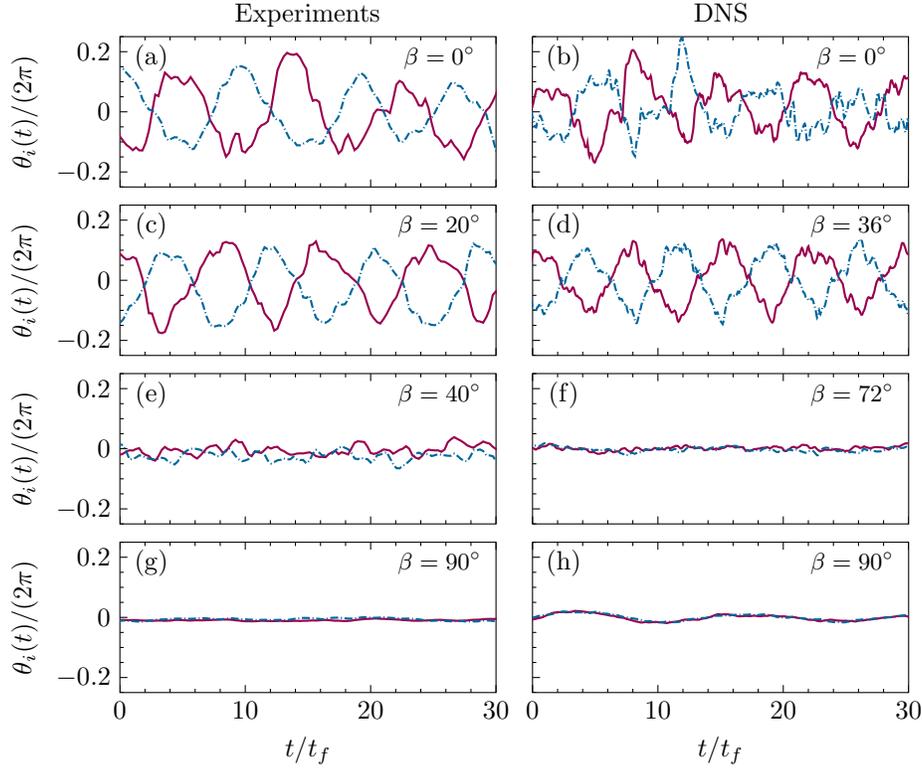}
\caption{Temporal evolution of the phase $\theta_1$ in the circle 1 (solid lines) 
and of the phase $\theta_5$ in the circle 5 (dash-dotted line) of the convection cell (see locations of the circles in figure~\ref{PIC2}),
as obtained in the experiments for  $\Ra=1.42\times 10^7$, $\Pran\approx0.0093$ $(a,\,c,\,e,\,g)$
and in the DNS for $\Ra=1.67\times10^7$, $\Pran=0.0094$ $(b,\,d,\,f,\,h)$
for the cell inclination angles $\beta=0^\circ$ $(a,\,b)$,
$\beta=20^\circ$ $(c)$, $\beta=36^\circ$ $(d)$, $\beta=40^\circ$ $(e)$, $\beta=72^\circ$ $(f)$
and $\beta=90^\circ$ $(g,\,h)$.}
\label{PIC8}
\end{figure}

In figure~\ref{PIC8} we present the dynamics of the LSC twisting and sloshing mode.
There,  the temporal evolution of the phases of the LSC in the circle 1 (closer to the heated plate) and in the circle 5 (closer to the cold plate) are presented 
for different inclinations angles $\beta$ of the convection cell filled with liquid sodium,
as it is obtained in our DNS and measurements.

The main evidence for the existence of the sloshing mode is the visible strong anticorrelation of the phases $\theta_1(t)$ and $\theta_5(t)$, which are measured via the probes at the circles 1 and 5, respectively.
It is present in the RBC case (figures~\ref{PIC8}~a,~b) and for the inclination angles $\beta=20^\circ$ (figure~\ref{PIC8}~c) and $\beta=36^\circ$ (figure~\ref{PIC8}~d).
The measurements and DNS at the inclination angles $\beta\geq40^\circ$ (figure~\ref{PIC8}~e) show that with the increasing $\beta$ the above anticorrelation vanish.
At large inclination angles, there is no visible anticorrelation of the phases $\theta_1(t)$ and $\theta_5(t)$ and one can conclude
that the sloshing movement of the LSC is not present anymore (figures~\ref{PIC8}~f,~g,~h).

The dynamics of the twisting and sloshing mode of the LSC can be further studied with the Fourier analysis.
Thus, from the DNS data ($\Ra=1.67\times10^7$, $\Pran=0.0094$) we obtain that the period duration $T_{s}(\beta)$ equals  $T_{s}(0^\circ)=6.9\,t_f$
for the RBC case and is equal to $T_{s}(36^\circ)=7.6\,t_f$ for the inclination angle $\beta=36^\circ$.
The experimental data give $T_{s}(0^\circ)\approx 9.2\,t_f$ for RBC and 
$T_{s}(20^\circ)\approx 8.7\,t_f$ for $\beta=20^\circ$.
Note that the frequency $\omega$ of the LSC twisting and sloshing is approximately proportional to the Reynolds number $\omega\cdot t_\kappa\sim \Rey$ \citep{Cioni1997}.
Therefore the slightly larger period duration in the experiments compared to those in the DNS are consistent with slightly lower Reynolds numbers and Rayleigh numbers there.
\begin{figure}
\centering
\includegraphics{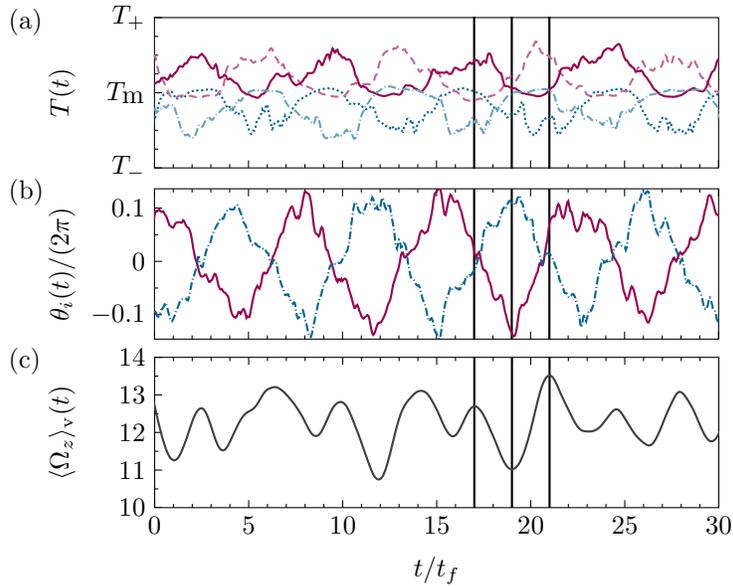}
\caption{Temporal evolution of different quantities obtained in the DNS
for $\Ra=1.67\times10^7$, $\Pran=0.0094$ and the cell inclination angle $\beta=36^\circ$:
$(a)$ the temperature $T$ at the probes B3 (solid line), D3 (dash-dotted line), F3 (dotted line) and H3 (dash line);
$(b)$ the phase $\theta_1$ in the circle 1 (solid line)
and the phase $\theta_5$ in the circle 5 (dash-dotted line) and
$(c)$ the volume-averaged component of the heat flux vector along the cylinder axis, $\avg{\Omega_z}{V}$.
The three vertical lines mark the times at which the snapshots in figure~\ref{PIC10} are taken.
The phases $\theta_1(t)$ and $\theta_5(t)$ in $(b)$ have a period of $T_\theta=7.6\,t_f$, which is determined by the Fourier analysis.}
\label{PIC9}
\end{figure}
\begin{figure}
\centering
\includegraphics{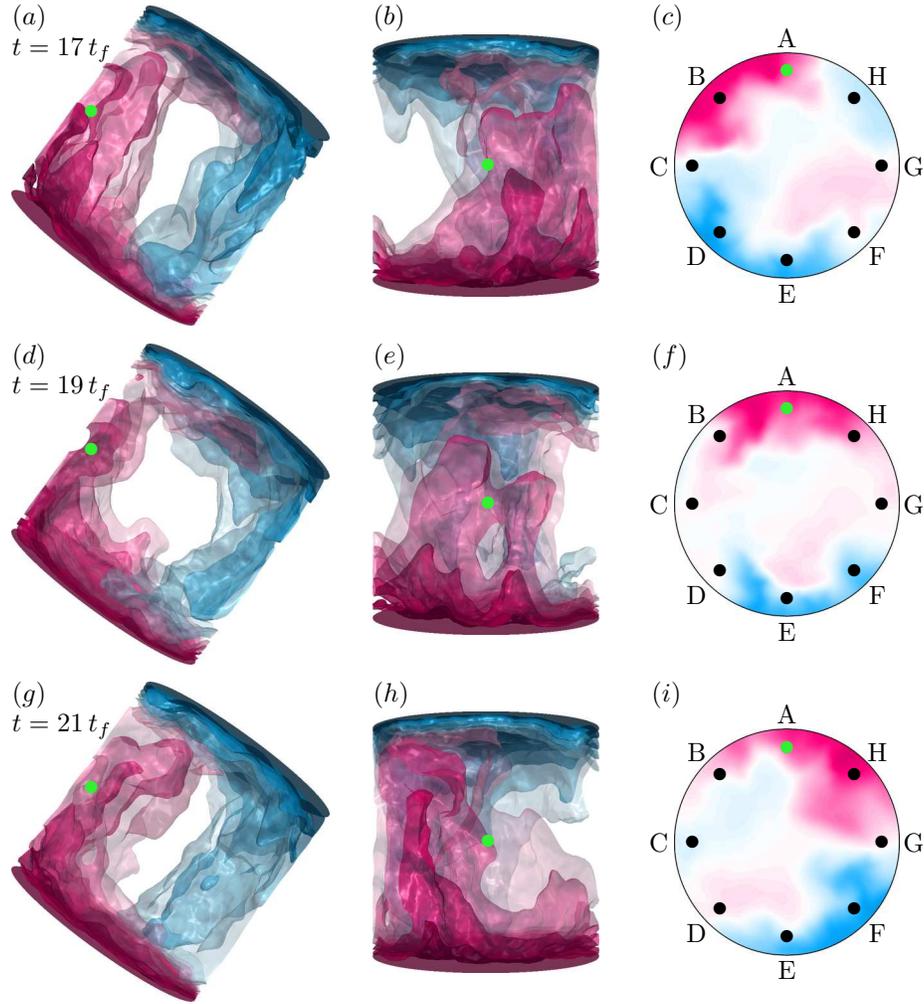}
\caption{3D side views $(a,\, d,\, g)$ and views orthogonal to the side views and to the cylinder axis $(b,\, e,\, h)$ of the temperature isosurfaces
and the corresponding horizontal slices at the mid-height of the instantaneous temperature fields as seen from the cold plate $(c,\, f,\, i)$,
which are obtained in the DNS for $\Ra=1.67\times10^7$, $\Pran=0.0094$ and the cell inclination angle $\beta=36^\circ$ at the times
$(a,\,b,\,c)$ $t=17\,t_f$, $(d,\,e,\,f)$ $t=19\,t_f$, $(g,\,h,\,i)$ $t=21\,t_f$.
The dot (green online) marks the location A3.
The here presented snapshots correspond to the times, marked in the figure~\ref{PIC9} with the vertical lines.}
\label{PIC10}
\end{figure}

Let us investigate the IC flow in liquid sodium for the inclination angle $\beta=36^\circ$, as in figure~\ref{PIC8}d,
where a very strong LSC sloshing is observed.
In figure~\ref{PIC9} we analyse this flow in more detail.
Figure~\ref{PIC9}a presents the evolution of the temperature in time, which is  measured by the probes
B3, D3, F3 and H3 that are placed in the central circle 3.
One can see that the temperature dependences on time at the locations B3 and F3 are perfectly synchronous.
So are the temperature dependences on time at the locations H3 and D3.
The temperatures at the locations B3 and D3 are anticorrelated. So are the temperatures at the locations H3 and F3.
Thus, when at the location B3 the fluid is extremely hot, the lowest temperature is obtained near D3, which is located only $90^\circ$ azimuthally below B3.
Analogously, when the fluid is hot at the location H3, its lowest temperature is obtained near the location F3, which is $90^\circ$ below H3
(see also figure~\ref{PIC10}).
These events happen at the times $t/t_f=17$ and $t/t_f=21$ in figure~\ref{PIC9}, respectively. 
Thus, at the times $t/t_f=17$ and $t/t_f=21$ a big hot and big cold plume approach each other very closely.
This sloshing movement happens periodically, alternately, near one side of the sidewall, then on the opposite side.

Figure~\ref{PIC9}b presents the evolution in time of the LSC phase $\theta_1$ in the circle 1 (close to the heated plate)
and of the phase $\theta_5$ in the circle 5 (close to the cold plate).
These phases are  anticorrelated and at the times when the sloshing brings together the hot and the cold streams of the LSC near the sidewall,
as described above (at $t/t_f=17$ and $t/t_f=21$), these phases are equal.
This means that at $t/t_f=17$ and $t/t_f=21$, there is no twisting of the LSC near the plates.
The twisting of the LSC near the plates is maximal (at $t/t_f=19$) 
when the hot and the cold streams of the LSC in the central cross-section are located near the opposite sides of the cylinder side wall.

The thermal plumes are emitted from the heated and cooled plates when the phase difference between $\theta_1$ and $\theta_5$ is maximal.
They travel towards the mid-plane and approach each other very closely when the  phase difference vanishes.
At each plate, the emission of the thermal plumes takes place, roughly speaking, from two different spots at the plate,
and again, this happens periodically and alternately.
Thus, the thermal plumes, while being emitted from the one spot, leave near the other spot sufficient space for new plumes to grow and detach from the thermal boundary layer.

In figure~\ref{PIC9}c the temporal evolution of the volume-averaged component of the heat flux vector along the cylinder axis, $\avg{\Omega_z}{V}$, is presented.
Again, a very strong relationship with the LSC phases and LSC sloshing is observed.
The maximal values of $\avg{\Omega_z}{V}$ are obtained when the hot and cold LSC streams meet, thanks sloshing,
while the minimum value is obtained at the time periods when the LSC is strongly twisted.

In figure~\ref{PIC10}, the above described process, namely, the azimuthal movement of the hot and cold batches of fluid in a form of an oscillatory motion against each other,
is illustrated with three-dimensional side views in two perpendicular directions.
Additionally, the corresponding horizontal cross-sections of the instantaneous temperature fields at the mid-height of the cylinder are presented there.
In the supplementary videos to this paper, the  described dynamics of the LSC can be observed in detail.
\begin{figure}
\centering
\includegraphics{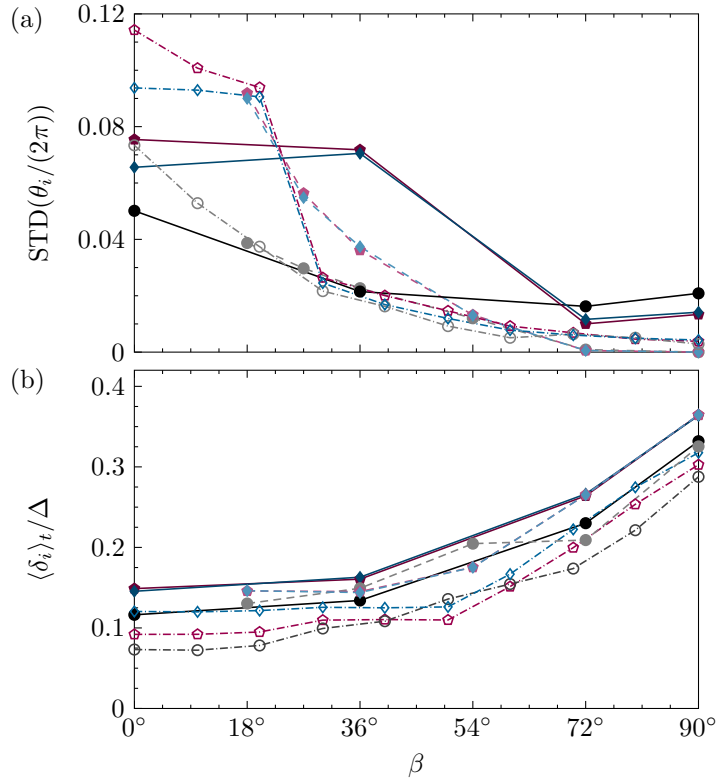}
\caption{$(a)$ Standard deviations of the phases $\theta_i$ and $(b)$ the time-averaged strengths of the large scale circulation, $\langle\delta_i\rangle_t$,
as obtained at the probe circle 1 (pink colour symbols and lines),
circle 3 (grey colour) and circle 5 (blue colour) in the experiments (dash-dotted lines), the DNS for $\Ra=1.67\times10^7$, $\Pran=0.0094$ (solid lines)
and DNS for $\Ra=10^6$, $\Pran=0.094$ (dash lines).}
\label{PIC11}
\end{figure}

In figure~\ref{PIC11}a the standard deviations of the phases $\theta_i$ in the circles $i=1,~3$ and 5 are presented, while
figure~\ref{PIC11}b shows the corresponding time-averaged strengths of the LSC, $\langle\delta_i\rangle_t$, as they are obtained in the liquid-sodium measurements and DNS.
The measurements show that the standard deviations of the phases $\theta_1$ (near the heated plate) and $\theta_5$ (near the cooled plate)
are relatively large for small inclination angles, while being small for large inclination angles.
There exist almost immediate drops of $\theta_1(\beta)$ and $\theta_5(\beta)$ that happen between $\beta=20^\circ$ and $\beta=40^\circ$,
which indicate a sharp transition between the twisting and sloshing mode of the LSC and usual mode of the LSC, when it is not twisted and 
located basically in the central vertical cross-section along the axis of the cylindrical sample.
The  standard deviations of $\theta_1$, $\theta_3$ and $\theta_5$, obtained in the DNS,   
show generally a similar behaviour as those measured in the experiments, but due to only a few considered inclination angles in the DNS,
it is impossible to resolve the sudden drop which is observed in the measurements.
Also one should notice that the data in figure~\ref{PIC11} are very sensitive to the time of statistical averaging,
which is extremely short in the DNS compared to the experiment.

The results for the time-averaged strengths of the LSC, $\langle\delta_i\rangle_t$,
obtained in the measurements and DNS (figure~\ref{PIC11}b) show good agreement.
In the RBC case ($\beta=0^\circ$), the LSC strength  is small and grows smoothly with the inclination angle $\beta$.
Surprisingly good here is the agreement between the liquid-sodium DNS data and the data 
from the auxiliary DNS for the same $\Ra\,\Pran$.

\subsection{Temperature and velocity profiles}

\begin{figure}
\centering
\includegraphics{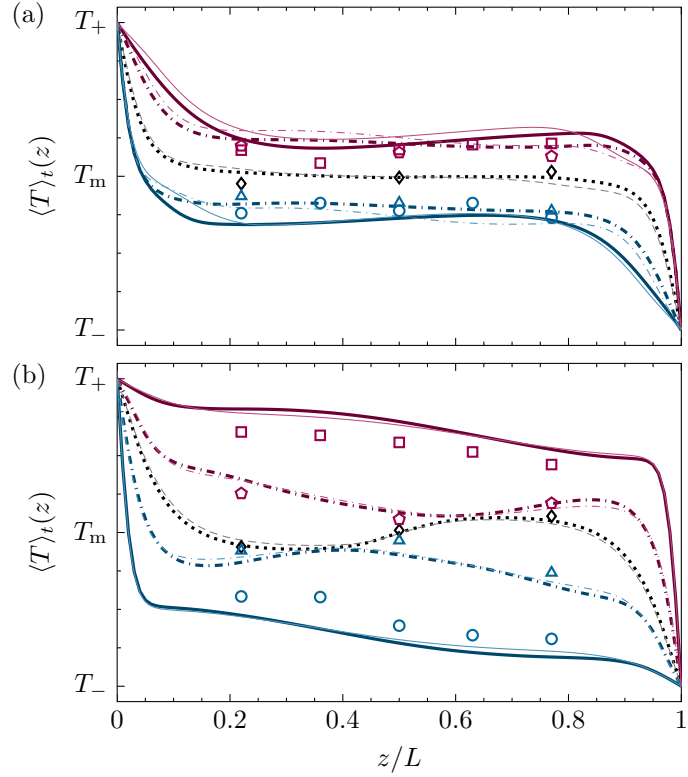}
\caption{Time-averaged temperature profiles at the positions A to H of the probes,
as obtained in $(a)$ the DNS for $\beta=36^\circ$ and the LES and experiments for $\beta=40^\circ$
and in $(b)$ the DNS for $\beta=72^\circ$ and the LES and experiments for $\beta=70^\circ$.
Thick lines are the DNS data, thin lines are the LES data and symbols are the experimental data.
Data at the position A (pink solid lines, squares) and the position E (blue solid lines, circles);
the average of the data at the positions B and H (pink dash-dotted lines, pentagons),
the average of the data at the positions D and F (blue dash-dotted lines, triangles)
and the average of the data at the positions C and G (black dotted and grey dash lines, diamonds).}
\label{PIC12}
\end{figure}
\begin{figure}
\centering
\includegraphics{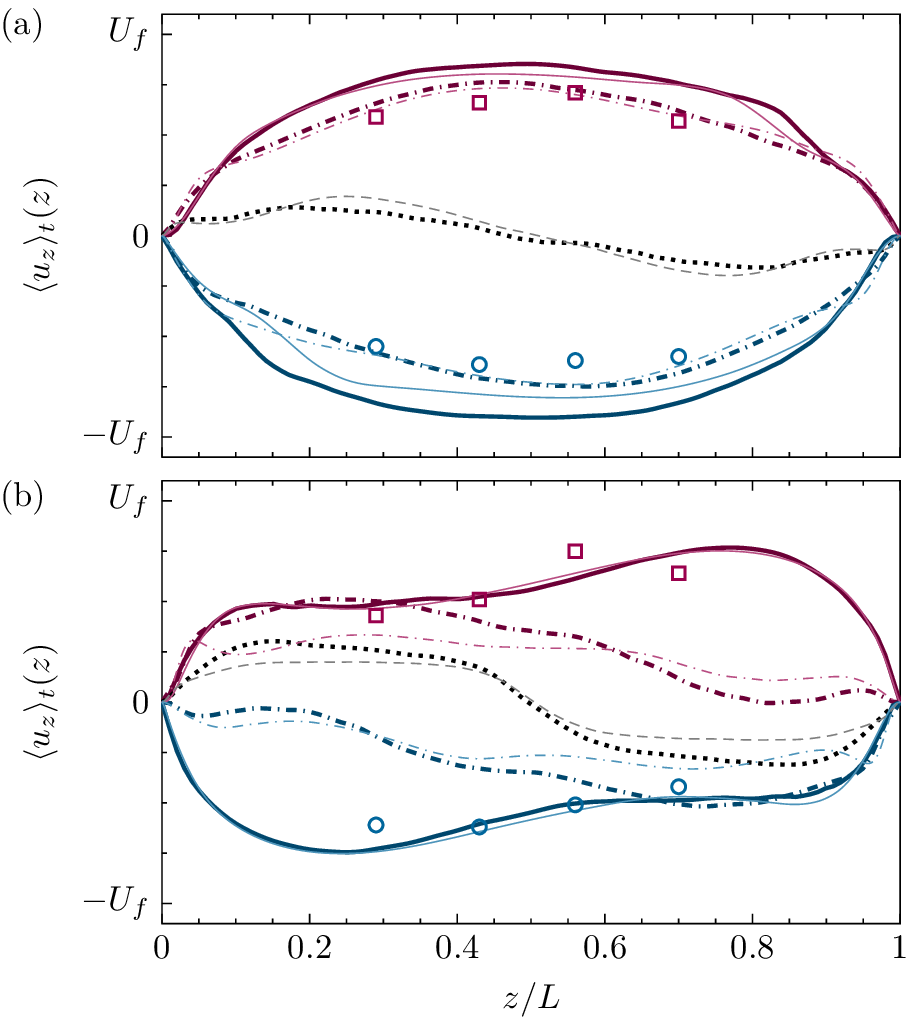}
\caption{Time-averaged profiles of the velocity component $u_z$, which is parallel to the cylinder axis, considered at the positions A to H of the probes,
as obtained in $(a)$ the DNS for $\beta=36^\circ$ and the LES and experiments for $\beta=40^\circ$
and in $(b)$ the DNS for $\beta=72^\circ$ and the LES and experiments for $\beta=70^\circ$.
Thick lines are the DNS data, thin lines are the LES data and symbols are the experimental data.
Data at the position A (pink solid lines, squares) and the position E (blue solid lines, circles);
the average of the data at the positions B and H (pink dash-dotted lines),
the average of the data at the positions D and F (blue dash-dotted lines)
and the average of the data at the positions C and G (black dotted and grey dash lines).}
\label{PIC13}
\end{figure}

In this section we analyse the temperature and velocity profiles.
The focus thereby is on the following two aspects. 
First, we compare the experimentally and numerically obtained profiles through the probes (positions A to H) along the lines aligned parallel to the cylinder axes. 
Second, we compare the velocity profiles, obtained in the DNS and LES,
with the velocities evaluated from the correlation times between two neighbouring probes in the experiment, 
in order to validate the method used in the experiment to estimate the Reynolds number.

In figure~\ref{PIC12}a, the time-averaged temperature profiles along the cylinder axis at the positions A to H are presented
for the inclination angles $\beta=36^\circ$ (DNS) and $\beta=40^\circ$ (LES and experiments).
Figure figure~\ref{PIC12}b shows analogous profiles 
for the inclination angles $\beta=72^\circ$ (DNS) and $\beta=70^\circ$ (LES and experiments).
In both figures, the profiles at the positions A and E are presented, as well as the average of the 
profiles at the positions B and H, the average of the profiles at the positions D and F
and the average of the C-profile and G-profile.
One can see that  for the same locations, the LES and DNS profiles are almost indistinguishable,
which again demonstrate excellent agreement between the DNS and LES.
The experimental data are available pointwise there, according to the 5  or 3 probes along each location, from  A to H.
The measurements are found to be also in good agreement with the numerical data, taking into account that the Rayleigh number in the experiments is about 15\% smaller than in the DNS.

By the inclination angle about $\beta=36^\circ$ or $\beta=40^\circ$ (figure~\ref{PIC12}a), the mean temperature gradient with respect to the direction $z$ across the plates
is close to zero in the core part of the domain.
This means that the turbulent mixing in this case is very efficient, which is also reflected in the increased Nusselt numbers that we studied before.
This efficient mixing is provided by the sloshing dynamics of the LSC.
In contrast to that, for the inclination angle about $\beta=70^\circ$ (figure~\ref{PIC12}b),
the mean flow is stratified and the temperature profiles have a non-vanishing gradients in the $z$-direction.

In figure~\ref{PIC13}, the time-averaged profiles along the cylinder axis of the velocity component $u_z$ are presented
for the same inclinations angles, as in  figure~\ref{PIC12}.
Again, a very good agreement between the DNS, LES and experiments is obtained.
The velocity estimates at the locations between the neighbouring thermocouples, which are
derived from the correlation times obtained in the temperature measurements,
are found to be in a very good agreement with the DNS and LES data.
Thus, this method to estimate the LSC velocity from the temperature measurement is proved to be a very reliable instrument in the IC liquid-sodium experiments.

\section{Conclusions}

In our complementary and cross-validating experimental and numerical studies,
we have investigated inclined turbulent thermal convection in liquid sodium ($\Pran\approx0.009$)
in a cylindrical container of the aspect ratio one.
The conducted measurements, DNS and LES demonstrated generally a very good agreement.
It was proved, in particular, that the usage of the cross-correlation time  
of the neighbouring temperature probes is a reliable tool to evaluate velocities during the temperature measurements in liquid sodium.

For the limiting cases of inclined convection, which are Rayleigh--B\'enard convection 
(with the cell inclination angle $\beta=0^\circ$) and vertical convection ($\beta=90^\circ$),
we have also studied experimentally the scaling relations of the mean heat flux (Nusselt number)
with the Rayleigh number, for $\Ra$ around $10^7$.
The scaling exponents were found to be about 0.22 in both cases, but the absolute values 
of $\Nu$ are found to be larger in VC, compared to those in RBC.
At the considered Rayleigh number about $1.5\times 10^7$,
any inclination of the RBC cell generally leads to an increase of the mean heat flux.
The maximal $\Nu$ is obtained, however, for a certain intermediate value of $\beta$.

For small inclination angles, the large-scale circulation exhibits a complex dynamics,
with twisting and sloshing. 
When the LSC is twisted, the volume-average vertical heat flux is minimal,
and it is maximal, when the LSC sloshing brings together the hot and cold streams of the LSC.
Figures~\ref{PIC9} and \ref{PIC10} and additional videos illustrate the studied LSC dynamics.
Additional investigations will be needed to study the even more complex behaviour of the LSC in IC of low-$\Pran$ fluids in elongated containers with $L\gg D$.

Furthermore we have found that for small Prandtl numbers there exist a similarity of the IC flows of the same $\Ra\,\Pran$,
due to the similar ratio between the thermal diffusion time scale, $t_\kappa$,  and the free-fall time scale, $t_f$,
for which holds $t_\kappa/t_f\sim\sqrt{\Ra\,\Pran}$.
Since in the small-$\Pran$ convective flows, the viscous diffusion time scale, $t_\nu$, is much larger than  $t_\kappa$,
the value of $\Ra\,\Pran$ determines basically the mean temperature and heat flux distributions.
The Nusselt numbers and the relative Reynolds numbers by inclination of the convection cell with respect to the gravity vector,
are also similar by similar values of $\Ra\,\Pran$.
This property can be very useful for the investigation of, e.g., the scaling relations of $\Nu$ and $\Rey$ with $\Ra$ and $\Pran$
or of the mean temperature or heat flux distributions,  by extremely high $\Ra$ and/or extremely small $\Pran$.

\subsubsection*{Acknowledgements}
This work is supported by the Priority Programme SPP 1881 Turbulent Superstructures of the Deutsche Forschungsgemeinschaft (DFG) under the grant Sh405/7.
O.S. also thanks the DFG for the support under the grant Sh405/4 -- Heisenberg fellowship.
The authors acknowledge the Leibniz Supercomputing Centre (LRZ) for providing computing time and
the Institute of Continuous Media Mechanics (ICMM UB RAS) for providing resources of the Triton supercomputer.

\bibliographystyle{plain}
\bibliography{ReferencesOlga}

\end{document}